\documentclass[onecolumn,preprint,12pt]{elsarticle}
\usepackage{graphicx, caption, subcaption}
\usepackage{epsfig}
\usepackage{amsfonts}
\usepackage{color}
\usepackage{ulem}
\usepackage{amsmath}
\usepackage{mathrsfs}
\usepackage[autostyle]{csquotes}
\usepackage{bigints}
\usepackage{floatrow}
\usepackage{caption}
\usepackage{enumitem}
\usepackage{etoolbox}
\usepackage{lineno}
\usepackage{hyperref}

\makeatletter
\patchcmd\frontmatter@PACS@format{\addvspace{11\p@}}{}{}{}% remove the space
\pretocmd\frontmatter@keys@format{\addvspace{11\p@}}{}{}
\patchcmd{\titleblock@produce}
  {\@pacs@produce\@pacs\@keywords@produce\@keywords}
  {\@keywords@produce\@keywords\@pacs@produce\@pacs}
  {}{}
\makeatother\makeatletter
\patchcmd\frontmatter@PACS@format{\addvspace{11\p@}}{}{}{}% remove the space
\pretocmd\frontmatter@keys@format{\addvspace{11\p@}}{}{}
\patchcmd{\titleblock@produce}
  {\@pacs@produce\@pacs\@keywords@produce\@keywords}
  {\@keywords@produce\@keywords\@pacs@produce\@pacs}
  {}{}
\makeatother

\usepackage{amssymb}
\journal{Physica A}

\begin{document}

\begin{frontmatter}



\title{The Impact of Foreign Players in the English Premier League: A Mathematical Analysis}

\author{A. Abdul}
 %\fntext[a]{}
  %\cortext[cor1]{}
\affiliation{                    
organization={Department of Applied Mathematics and Data Science,
Aston Centre for Artificial Intelligence and Research Applications (ACAIRA), Aston University}, addressline={Aston Triangle}, city={Birmingham}, \postcode={B4 7ET}, country = {United Kingdom}}
%\email{a.k.chattopadhyay@aston.ac.uk}

\author{Amit K Chattopadhyay\corref{cor1}}
\cortext[cor1]{Corresponding Author}
%\fntext[Corresponding author]{Corresponding author}
\affiliation{                                  
organization={Department of Applied Mathematics and Data Science,
Aston Centre for Artificial Intelligence and Research Applications (ACAIRA), Aston University}, addressline={Aston Triangle}, city={Birmingham}, \postcode={B4 7ET}, country = {United Kingdom}}
\ead{a.k.chattopadhyay@aston.ac.uk}
\ead[url]{https://research.aston.ac.uk/en/persons/amit-chattopadhyay}

\author{S. Jain}
\affiliation{                                  
organization={Department of Applied Mathematics and Data Science,
Aston Centre for Artificial Intelligence and Research Applications (ACAIRA), Aston University}, addressline={Aston Triangle}, city={Birmingham}, \postcode={B4 7ET}, country = {United Kingdom}}
\ead{s.jain@aston.ac.uk}

\begin{abstract}
We undertake extensive analysis of English Premier League data over the period 2009/10 to 2017/18 to identify and rank key factors affecting the economic and footballing performances of the teams. Alternative end-of-season league tables are generated by re-ranking the teams based on five different descriptors - total expenditure, total funds spent on players, total funds spent on foreign players, the ratio of foreign to British players and the overall profit. The unequal distribution of resources and expenditure between the clubs is analyzed through Lorenz curves. A comparative analysis of the differences between the alternative tables and the conventional end-of-season league table establishes the most likely factors to influence the performances of the teams that we also rank using Principal Component Analysis. We find that the top teams in the league are also those that tend to have the highest expenditure overall, for all players, including foreign players; they also have the highest ratios of foreign to British players. Our statistical and machine learning study also indicates that successful performance on the field may not guarantee healthy profits at the end of the season.
\end{abstract}

\begin{graphicalabstract}
\includegraphics[height=0.75\textheight,width=1.2\textwidth]{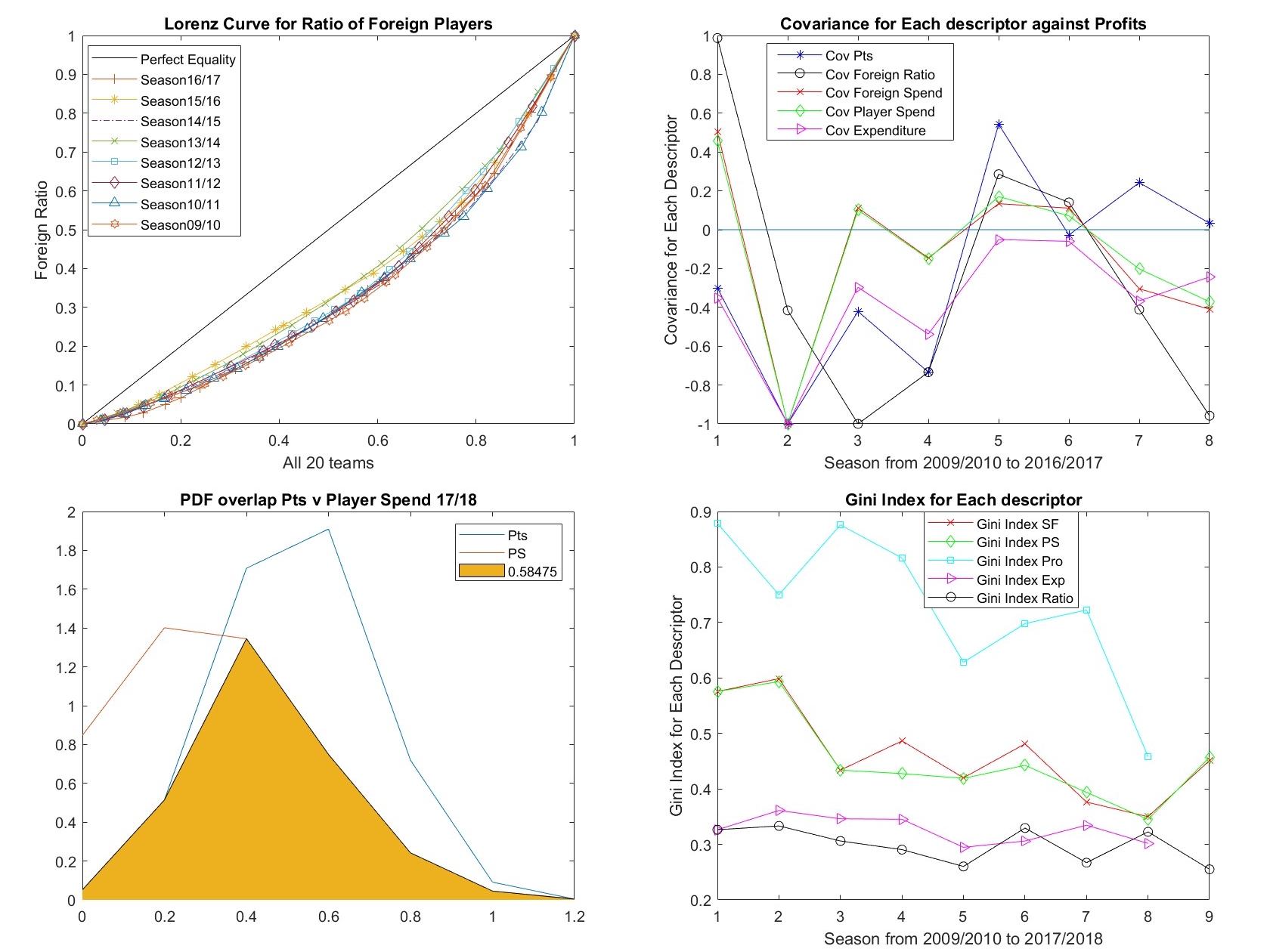}
\end{graphicalabstract}

\begin{highlights}
\item Data modelling to predict and quantify the impact of foreign players in soccer leagues
\item Machine Learning-based performance rating in soccer and related econometrics
\item On-field performance may not statistically translate into economic profit in soccer
\end{highlights}

\begin{keyword}
English Premier League \sep
Correlation coefficients \sep
Lorenz curve \sep
PDF \sep
\PACS 42.50.Ex \sep 32.80.Wr \sep 32.80.-t \sep 32.10.Fn
\end{keyword}

\end{frontmatter}

\section{Introduction}
\label{intro}
\begin{figure}[H]
	\centering
	\includegraphics[width=0.5\textwidth]{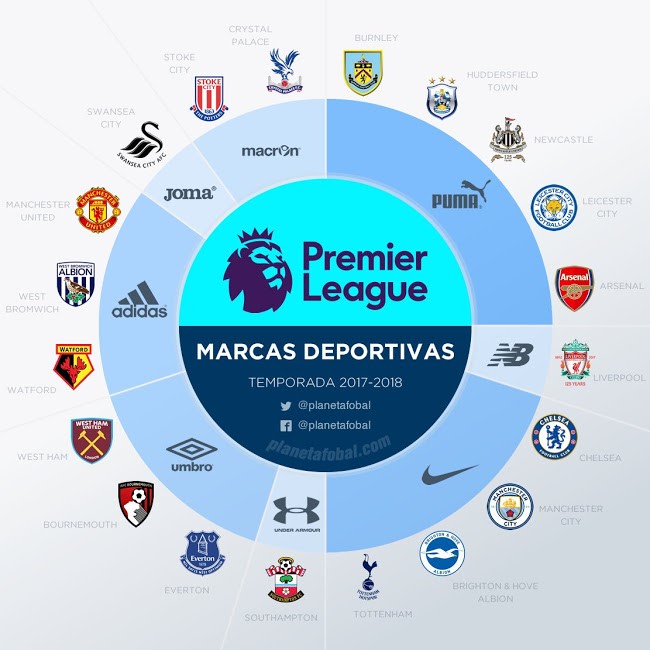}
	\centering
	\caption{An illustration of some of the different sponsors of teams in the Premier League for the 2016/17 season}
	\label{label:fig1}
\end{figure}

The English Premier League (EPL) was founded in 1992 and, over time, has built up a reputation for being one of the most competitive leagues in European soccer \cite{Pinnacle}. As a consequence, the EPL has attracted numerous sponsors who wish to benefit from its global reach and appeal, for example through kit sponsorship shown in Figure \ref{label:fig1}. The sponsorship and, in addition, the increased funding resulting from broadcasting rights, have permitted the clubs to invest in the best players and other assets. As is well known, 
the \lq better\rq\ perceived players tend to cost more.

Considerable financial activity surrounds the movement of players (and, hence, funds) between clubs during the two annual transfer windows. The underlying motivation behind the activity is usually either to reinforce a club's challenge at the top of the table or a rearguard action in the dangerous relegation zone \cite{SkySports}. The transfer of players takes place in a global market. As a consequence, the composition of the EPL teams we see today is very different to what it was when the league was first launched in 1992. Indeed, it has been described as the Premier League's worldwide pool of talent \cite{mail}. As far back as the 2012/13 season, only 32 Premier League starts were by players qualified to play for England, compared to 69 twenty years ago \cite{FACom}. Nowadays, the EPL has one of the highest percentages of foreign players when compared to other European football leagues at 69.2 \cite{Sky} and 65 different nationalities are represented in the competition with French and Spanish being the most common \cite{Sky}. Data analysis has played a key role in this as clubs are now able to draw up a shortlist of players whose playing statistics match the profile of their ideal target signing \cite{ft}. The low percentage of players in the EPL eligible to play for England led the former FA Chairman Greg Dyke to call a commission to address the issue as it was felt that there could be consequences if the selection pool for the national team contained insufficient numbers of players participating in the sport at the highest level.

First, we explain how performance in the football field is quantified in the EPL structure. Points determine the position of a team in the EPL table. Three points are awarded for a win, one point for a draw and none for a loss. At the end of a season, the total number of points for each team determines the winner of the league and also those to be relegated. The three teams with the fewest points are relegated to the Championship; their places are taken by three teams promoted from the Championship. If the number of total points is the same between two teams, then the teams are ranked by goal difference which is the number of goals scored minus the number of goals conceded. The higher the goal difference, the higher the ranking of the team. If the rank of the two teams still cannot be separated by total points or goal difference, then goals scored by the teams are used to separate them, with the team with more goals ranking higher. If that cannot separate the teams, then they receive the same rank. There are 20 teams in the EPL and they play each other twice in a season from August to May, one home fixture and one away. As a result, each team plays 38 games in the full season.

The main purpose of the present work is to examine the impact of foreign players on the EPL. We want to investigate whether foreign players improve the performance of their clubs and/or generate sufficient additional income to justify the initial expenditure.
It has been argued that the uncertain outcome of investment in players, influences the amount of revenue that the club can generate and therefore the capacity to recoup the investment cost \cite{MoneyPaper}. 

Although the main indicator of a team's success over a given season is the club's position in the end-of-season league table, numerous studies have been undertaken in an attempt to quantify the full or true measure of achievement \cite{Guard}. Oberstone \cite{DiffPaper}, for example, has developed a regression model to study five independent criteria in addition to the end-of-season points in an attempt to differentiate the various teams at the top of the table: individual performance of the players on the pitch; quality of defence; number of goal attempts; discipline of players; and the consistency of passes made during matches. Each individual descriptor was further refined in order to extract additional information. Somewhat unsurprisingly, the conclusion was that the success of a team was most likely to be determined by the collective performance of its players throughout the season. A key variable in this model relates to the quality of the defence of a team. This observation is supported by the belief of many commentators that the number of goals conceded during a season by a team is probably one of the most important aspects \cite{ESPN} that should be considered. The importance of an organized and effective defence is also discussed by \cite{TpPaper} in a rather limited study restricted to just a single season's data \cite{DetPaper}. Hence, there is growing quantitative evidence that a team's overall performance is determined by the contributions of individual players on the field.

The significance of the international transfer market and its effects on the EPL were studied by Madichie \cite{Nna} who argued that foreign players have contributed positively to the development of top league football teams. However, Madichie \cite{Nna} refers only to high-profile players who have an additional marketing appeal which can make a substantial financial contribution to a club's balance sheet \cite{Hand}. For example, the purchase of Cristiano Ronaldo by Juventus for over \$ 100 million enabled his new club to sell \$ 60 million worth of his jerseys within 24 hours \cite{cnbc} in 2018.

Other authors have attempted to re-rank the EPL table based on new criteria in order to obtain a league table which captures additional relevant information. Firth \cite{DF}\ re-ranked the table by incorporating the schedule strength of each team. The resulting table takes into account the opportunity to gain points by considering whether a team is playing at home or away and also the difficulty of the matches played.

Several authors have also studied related problems in football. For example, an empirical study of the distribution of goals in football leagues in Italy, England, Spain and Brazil was undertaken by Malacarne and Mendes \cite{MM}. Surprisingly, they found distributions not too dissimilar to those that emerge within non-extensive statistical mechanics. Hidden power laws have also been discovered in the European football leagues \cite{SMS} as well as a power law distribution for the tenure length of sports managers \cite{ALSS}.

In the next section, we describe the methodology used in our work. We first discuss the data and then the re-ranking technique used to re-rank the teams according to five different criteria employed. This is followed by an analysis of the distributions of the resources and expenditures throughout the EPL; each criterion is discussed in detail. In the subsequent section, we discuss the results from a multivariate investigation of the data. We conclude with a summary of our findings, especially those relating to the impact of foreign players in the EPL.
\section{Methods}
\subsection{Problem Statement}
In this article, we re-rank the final EPL table from 2010 to 2017 using five different criteria: proportion of foreign to British players; expenditure on foreign players; expenditure on all players; overall profits; and overall expenditure. This re-ranked structure is then compared against the official rankings based purely on the performance of the teams throughout the season. Our objective is to statistically analyze key optimization factors balancing and affecting sporting performance against the econometrics involved in purchasing (or selling) players as well as other monetary incentives like television rights and advertisements. As a result, we specifically address the question: \lq Do teams with a higher proportion of foreign to British payers outperform those with a lower figure?\rq\ An associated question to this is if teams spending more on all players outperform those spending less. We then focus on the key question of monitoring the expenditure of foreign players. Is it statistically true that teams spending more on foreign players outperform those spending less, or in a similar vein, do teams with more profits outperform less profitable ones? Overall, how does total expenditure contribute to the sporting performance graph? What we do not explicitly address are auxiliary questions like the sustainability of the expenditure graphics and the overall loss of local face in the money-dominated leagues. 

The above questions are investigated statistically to answer how each criterion correlates with the performance of the teams as measured by the conventional league table. The unequal distribution of the players, wealth and expenditure amongst the clubs is then measured via Gini coefficients or, equivalently, Lorenz curves \cite{Lor} and also by evaluating the overlaps between probability distribution curves. Finally, we carry out a principal component analysis.
\subsection{Data Collection}
The empirical data analyzed in this work were acquired from a number of football and general sports-related websites over a period from 2010 to 2017. Although the information is available in tabulated form on the web pages, it had to be parsed into Matlab for further reformatting and manipulation. Data for the conventional end-of-season EPL standings were obtained from \cite{Whats}. Information regarding the other criteria of interest in the present study was extracted from various different sources. For example, the nationalities of the players in each squad were downloaded from \cite{Whats} and the price paid (in millions of pounds) by the teams for players during the transfer windows from \cite{SoccerNews}. Profit and expenditure details for the teams were obtained from the data held by Companies House \cite{CompaniesHouse}.
\subsection{Re-ranking the English Premier League}
As mentioned above, we employed five different criteria to re-rank the league tables. We will use the data at the end of the 2009/10 season to illustrate our procedure. Table \ref{table:1} below contains 13 columns of data from 2009/10. The ones in the middle (columns 6-8) contain the conventional information as found in a standard EPL table at the end of the season. Columns 9-13 contain the data as discussed below for our 5 different criteria. Finally, in columns 1-5 we re-rank the teams using the additional data. We discuss each of the criteria below in detail before performing the re-ranking process.
\subsection{Ratio of foreign to British players}
Here we are interested in the ratio of foreign players to those with British nationality (English, Scottish, Welsh and Northern Irish) in each squad. For example, this is shown in column 9 of Table \ref{table:1} for the 2009/10 season. The teams are then re-ranked using this data in column 5 in order of an increasing ratio i.e. the team with the lowest ratio (Birmingham) is ranked first.
\subsection{Total expenditure on all players}
The total expenditure on all players refers to the spending undertaken by the clubs for players during the two annual transfer windows. This is displayed in column 10 of Table \ref{table:1} for 2009/10. The teams are then re-ranked in column 3 in order of increased spending, with the lower-spending teams at the top. One would expect to see the teams that have spent the most in transfer windows to be also placed the highest according to the official rank based on points (columns 6 and 8) as financial performance has become one of the key features of football \cite{Hand}.
\subsection{Total expenditure on foreign players}
The total expenditure on all foreign players purchased by the clubs is obviously a subset of the expenditure on all players. This is shown in column 11 for the 2009/10 season. The re-ranking of the teams is performed in order of increased spending and is displayed in column 4. Once again, one expects to see teams that have spent the most on foreign players achieve more points in a season and, thereby, finish higher up in the official league table. This is clearly a consequence of wealthier clubs being able to purchase high-profile players instead of having to develop talent \cite{Solberg}.   
\subsection{Total expenditure}
The total amount of expenditure by teams in the EPL each season includes all employee costs (players, groundsman, maintenance, etc.), auditor's remuneration, tangible and intangible assets and investments; this is displayed in column 13. In our re-ranking scheme, we rank the teams in order of increasing expenditure, with the team spending the least at the top. This is shown in column 1 of Table \ref{table:1}.
\subsection{Total profit}
The total profit for each team for a season is the difference between the total income and the total expenditure. We display this in column 12 and the re-ranked data in column 2 where we rank the teams in order of increasing profit, with the team making the smallest profit at the top.
\newline

\begin{table}[ht]
\centering
\resizebox{\linewidth}{!}{
\begin{tabular}{|c|c|c|c|c|c|c|c|c|c|c|c|c|}
\hline
 Expense & Profit & Total & Foreign & Foreign: & Position & Team & Pts & Ratio & Total  & Foreign & Profits & Exp\\ 
 (\pounds m) & (\pounds m) & spend (\pounds m) & spend (\pounds m) & British  &  & &  &  & spend (\pounds m) & spend (\pounds m) & (\pounds m) & (\pounds m) \\
 %\midrule
\hline 
20 & 2 & 12 & 14 & 17 & 1 & Chelsea & 86 & 2.4444 & 21.8 & 21.8 & -70.437 & 257.727 \\ 
17 & 18 & 13 & 15 & 10 & 2 & Man U & 85 & 1.1111 & 22 & 22 & 13.544 & 191.568 \\ 
16 & 20 & 10 & 10 & 18 & 3 & Arsenal & 75 & 3.25 & 11.2 & 11.2 & 92.32 & 179.496 \\ 
15 & 10 & 14 & 11 & 6 & 4 & Tottenham & 70 & 0.8 & 22.6 & 12.1 & -5.163 & 134.517 \\ 
19 & 1 & 20 & 20 & 14 & 5 & Man C & 67 & 1.5833 & 89.3 & 61.8 & -117.793 & 253.801 \\ 
14 & 4 & 18 & 13 & 3 & 6 & Aston V & 64 & 0.66667 & 35.6 & 21.6 & -27.712 & 117.198 \\ 
18 & 7 & 19 & 17 & 19 & 7 & Liverpool & 63 & 3.4286 & 44.8 & 23.3 & -19.935 & 227.683 \\ 
13 & 12 & 15 & 16 & 7 & 8 & Everton & 61 & 0.9375 & 23 & 23 & -3.093 & 101.26 \\ 
5 & 16 & 5 & 1 & 1 & 9 & Birmingham & 50 & 0.52632 & 3.4 & 0 & 0.199 & 56.515 \\ 
8 & 13 & 2 & 4 & 20 & 10 & Blackburn & 50 & 3.5 & 2.3 & 2.3 & -1.896 & 70.425 \\ 
7 & 11 & 11 & 12 & 8 & 11 & Stoke & 47 & 1 & 18 & 14.5 & -4.517 & 66.532 \\ 
11 & 8 & 7 & 7 & 13 & 12 & Fullham & 46 & 1.5 & 4.7 & 4.7 & -16.942 & 97.02 \\ 
12 & 5 & 17 & 18 & 4 & 13 & Sunderland & 44 & 0.6875 & 33.3 & 26.3 & -26.179 & 97.149 \\ 
9 & 3 & 16 & 19 & 12 & 14 & Bolton & 39 & 1.4615 & 27.1 & 27.1 & -35.443 & 90.461 \\ 
4 & 19 & 9 & 9 & 5 & 15 & Wolvs & 38 & 0.77778 & 10.5 & 10.5 & 16.29 & 44.354 \\ 
1 & 15 & 4 & 6 & 16 & 16 & Wigan & 36 & 2 & 2.9 & 2.9 & 0.075 & 3.677 \\ 
10 & 6 & 3 & 5 & 11 & 17 & West Ham & 35 & 1.1429 & 2.5 & 2.5 & -21.485 & 94.262 \\ 
3 & 17 & 6 & 2 & 2 & 18 & Burnley & 30 & 0.52632 & 3.5 & 0 & 10.247 & 40.372 \\ 
6 & 9 & 8 & 8 & 9 & 19 & Hull & 30 & 1 & 5.8 & 5.8 & -6.831 & 58.154 \\ 
2 & 14 & 1 & 3 & 15 & 20 & Portsmouth & 19 & 1.8 & 2 & 0 & 0 & 5.4 \\ 
\hline 
\end{tabular}
}
\caption{Re-rank table for 2009/10 Season. Total Player Spend, Total Foreign Spend, Profits and Expenditure are in millions of pounds. Negative profit means a loss.}
\label{table:1}
\end{table}

\noindent The complete data for the re-ranked tables for the eight seasons from 2009/10 to 2016/17 can be found in Appendix 1.
There are clear trends to be discerned from these tables: teams finishing higher up in the EPL at the end of a season tend to be at the lower end of all of our re-ranked tables apart from the one relating to the total profit. Hence, the top teams in the EPL are also those that have the highest total expenditure for the three criteria studied (all players, foreign players and overall), as well as the highest ratio of foreign to British players. An obvious exception to our overall observation occurred in 2015/16 when Leicester City were crowned the winners against all odds. We explore this point in more detail later.

\par
\begin{figure}[H]
	\centering
	\includegraphics[width=\textwidth]{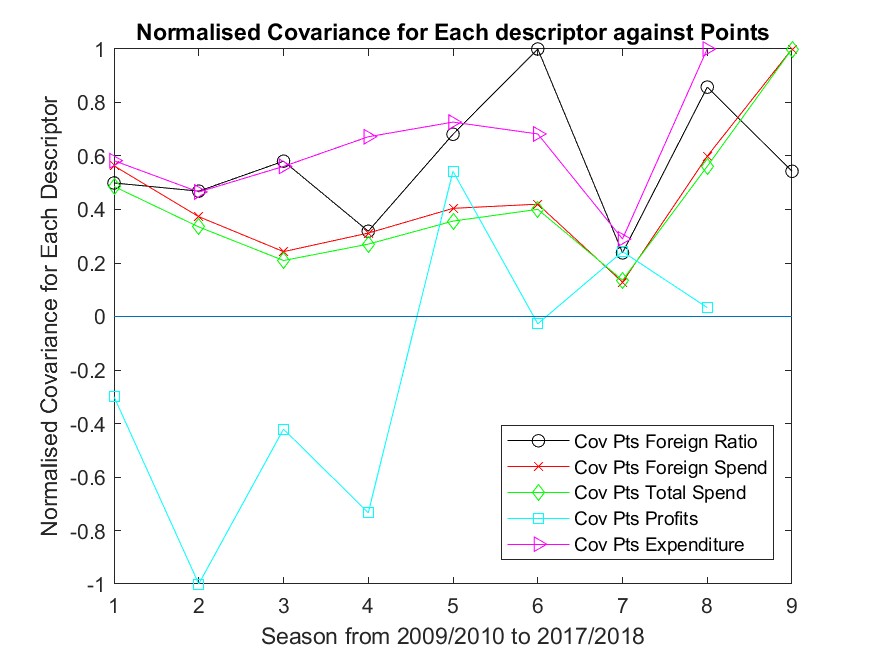}
	\centering
\caption{Correlation coefficients for each descriptor against the number of points. Note that although complete data is displayed for the seasons from 2009/10 to 2016/17, we have only partial data for 2017/18; the data for profits and expenditure for this season was unavailable when this study was undertaken.}
\label{label:fig2}
\end{figure}

In Figure \ref{label:fig2}, we plot the correlation coefficients for each descriptor (our criterion) against the number of points obtained for all of the seasons studied. It is evident that all of the descriptors, apart from the total profit, are correlated with each other. 

\par
In the next section, we discuss the results of an extensive analysis of the data.

\section{Results and Discussions}
\subsection{Measures of Inequality}
The funds available to the teams in the EPL are not distributed evenly and, therefore, neither is the expenditure of the clubs. We estimate the unequal distribution of the resources and the performances of the teams by evaluating several different measures of inequality. The robustness of any given measure should be confirmed by an alternative technique.
%\subsection{The Lorenz Curve}
In order to measure the inequality of the distribution of our five descriptors we first evaluated the relevant Gini coefficients and Lorenz curves for the data \cite{Lor}. 
\begin{figure}[H]
	\centering
	\includegraphics[width=\textwidth]{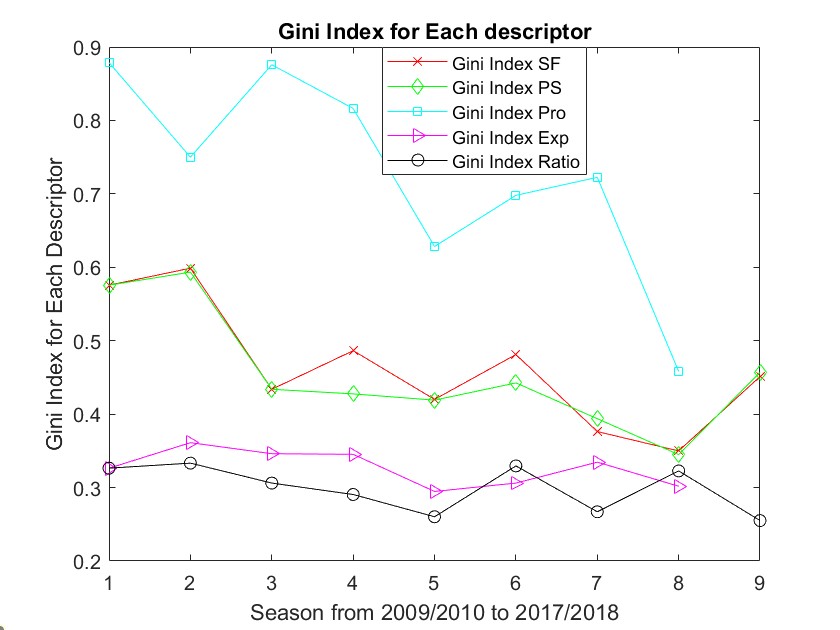}
	\centering
%\end{subfigure}
\caption{Gini indices for all 5 descriptors spanning years 2009/10 to 2017/18.}
\label{label:fig_Gini}
\end{figure}
In all figures, we have consistently used circles to indicate \enquote{ratio of foreign to British players}, crosses to indicate \enquote{spending on foreign players}, inverse ellipses to indicate \enquote{total spend}, squares to indicate \enquote{total economic profit} and forward triangles to indicate \enquote{total expenditure}. As can be clearly seen from Figure \ref{label:fig_Gini}, the profit margin shows the highest values with maximum unevenness in distribution over the years while the ratio of foreign to British players (stipulated to 4 maximum) is the most steady profile. The greater inequality in the profit margin is not unexpected though, especially in the context of investment in foreign players that is often much larger than that for the local players.
\begin{figure}[H]
\center
%\begin{subfigure}[b]{0.53\textwidth}
	\centering
	\includegraphics[width=\textwidth]{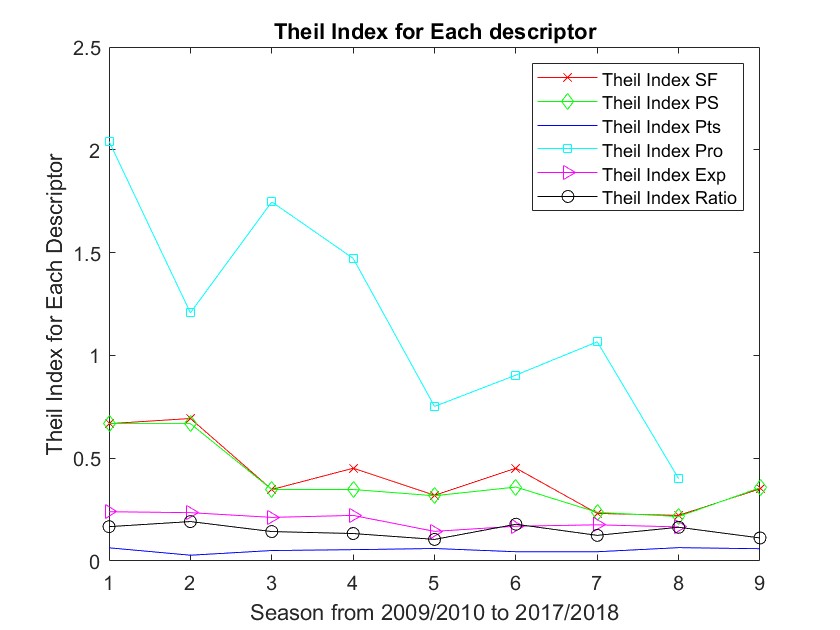}
	\centering
\caption{Theil indices for all 5 descriptors spanning years 2009/10 to 2017/18.}
\label{label:fig_Theil}
\end{figure}
Figure \ref{label:fig_Theil} shows comparative Theil indices for the same descriptor data as in Figure \ref{label:fig_Gini}. Clearly, both inequality indices, Gini and Theil, mostly track the inequality baseline but the absolute values are different. Due to the multivariate nature of the data and their interdependence, we tend to rely more on the inequality information from the Theil indices than from Gini indices. 

\begin{figure}[H]
\begin{subfigure}[b]{0.48\textwidth}
	\centering
	\includegraphics[width=\textwidth]{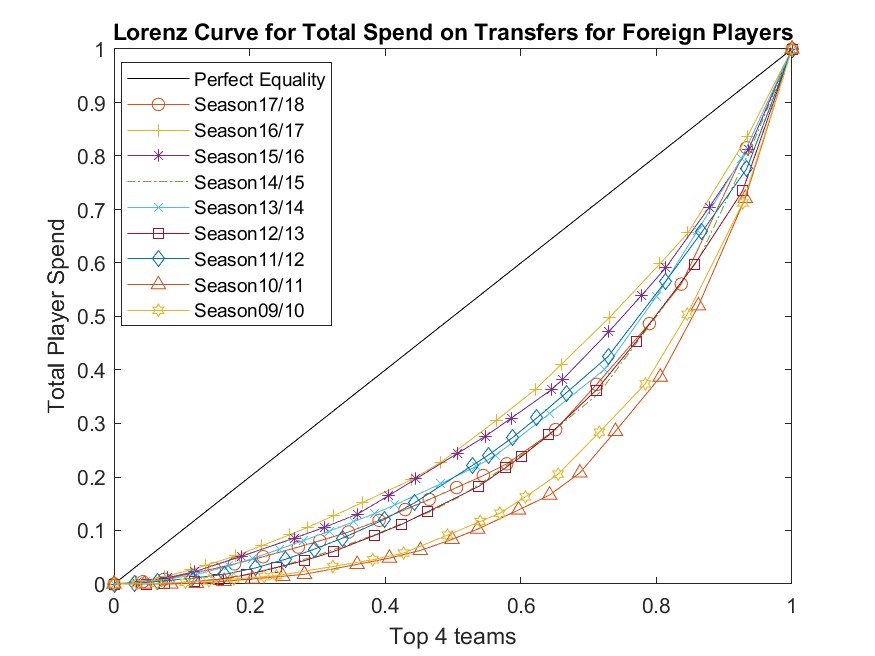}
	\caption{\centering Lorenz Curves for Total Player Spend on Foreign Players}
	\centering
	%\label{label:file_name}
\end{subfigure}
\hfill
\begin{subfigure}[b]{0.48\textwidth}
	\centering
	\includegraphics[width=\textwidth]{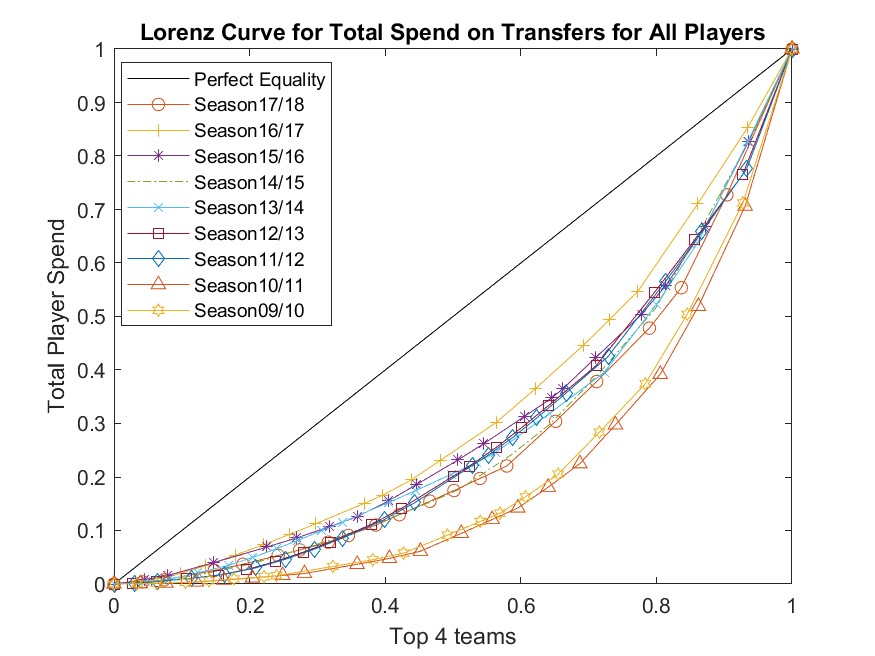}
	\caption{\centering Lorenz Curves for Total Player Spend on All Players}
	\centering
	%\label{label:file_name}
\end{subfigure}
\center
\begin{subfigure}[b]{0.48\textwidth}
	\centering
	\includegraphics[width=\textwidth]{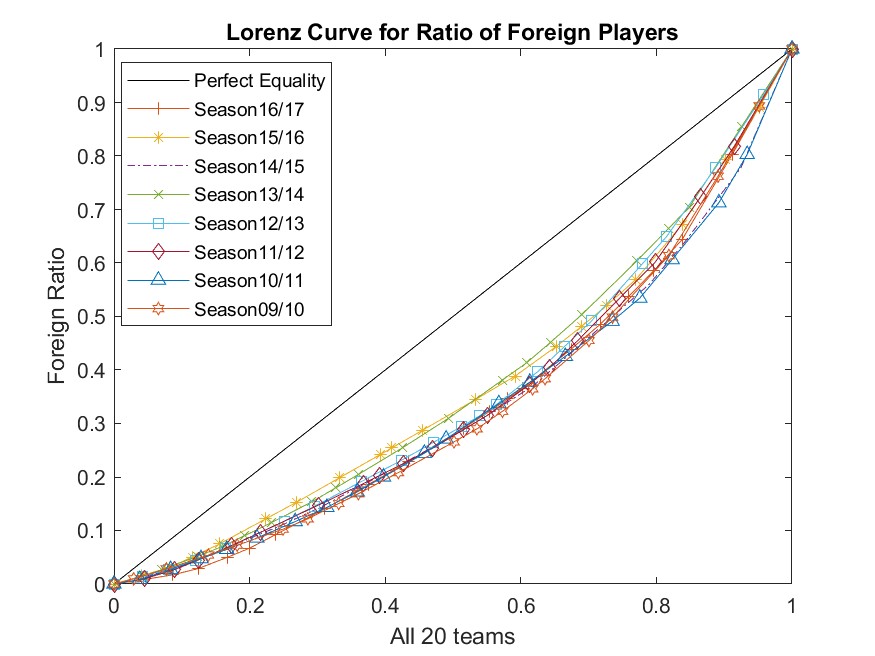}
	\caption{\centering Lorenz Curves for Ratio of Foreign to British Players}
	\centering
	%\label{label:file_name}
	\end{subfigure}
	
	\begin{subfigure}[b]{0.48\textwidth}
	\centering
	\includegraphics[width=\textwidth]{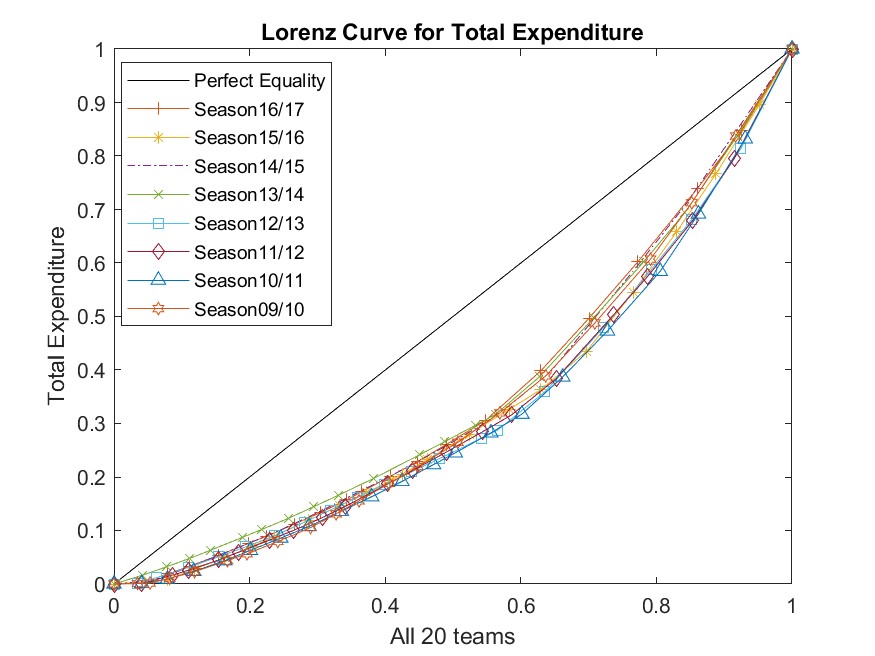}
	\caption{Lorenz Curves for Expenditure}
	\centering
	%\label{label:file_name}
\end{subfigure}
\hfill
\begin{subfigure}[b]{0.48\textwidth}
	\centering
	\includegraphics[width=\textwidth]{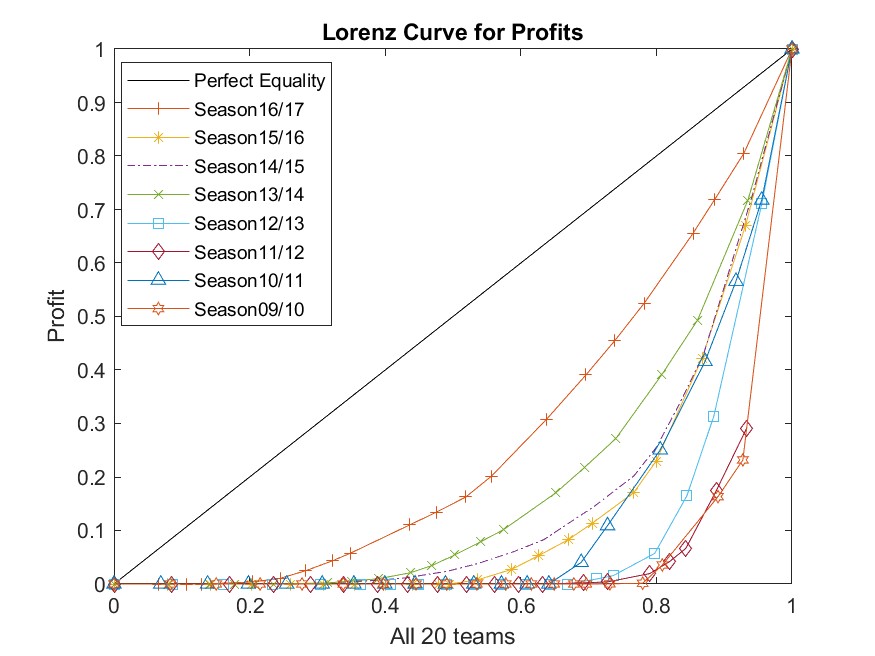}
	\caption{\centering Lorenz Curves for Profits}
	\centering
	%\label{label:file_name}
\end{subfigure}
\caption{Lorenz curves for each descriptor against points}
\label{label:fig_Lorenz}
\end{figure}

The curves for all five descriptors for all of the seasons investigated are shown in Figures \ref{label:fig_Lorenz}(a-e). Note that curves \ref{label:fig_Lorenz}(a) and \ref{label:fig_Lorenz}(b) display additional data for the 2017/18 season. The deviation from the line of perfect equality (the straight line at $45^o$ in each plot) is an indication of the amount of unequal distribution of the quantity of interest. We remind the reader that the Gini coefficient in any particular scenario is simply the ratio of the area between the line of equality and a given curve to the total area under the straight line.

%\newpage

%\section{Discussion}
\subsection{Probability Distribution Function Overlap}
The overlapping area of any two given probability distribution curves is an estimate of the agreement of the two descriptors concerned. The numerical value can range from zero (total disagreement) to one (total agreement).
In Table \ref{table:2} we display the values for the non-overlapping areas of the probability distribution curves for the points and the five descriptors of interest. (Note: non-overlapping areas can range from zero (total agreement) to one (total disagreement)).

\begin{table}[ht]
\centering
\resizebox{\linewidth}{!}{%
\begin{tabular}{|c|c|c|c|c|c|c|c|c|c|c|c|c|}
%\toprule
\hline
\textbf{Season} & \textbf{Pts v Ratio} & \textbf{Pts v Player Spend} & \textbf{Pts v Foreign Spend} & \textbf{Pts v Profits} & \textbf{Pts v Expenditure} \\ 
 \hline
%2017/18 & 0.1878 & 0.4771 & 0.4439 &  NA &  NA\\
\hline
2016/17 & 0.4451 & 0.4554 & 0.4485 &  0.6346 &  0.3995\\
\hline
2015/16 & 0.4829 & 0.7280 & 0.6474 & 0.8905 & 0.5191\\
\hline
2014/15 & 0.7037 & 0.6196 & 0.5775 & 0.7307 & 0.4058\\
\hline
2013/14 & 0.4259 & 0.5958 & 0.6023 & 0.7062 & 0.4762\\
\hline
2012/13 & 0.3782 & 0.6538 & 0.6918 & 0.8838 & 0.5543\\
\hline
2011/12 & 0.4312 & 0.5549 & 0.6599 & 0.8703 & 0.6657\\
\hline
2010/11 & 0.7679 & 0.8221 & 0.8108 & 0.9484 & 0.6827\\
\hline
2009/10 & 0.4208 & 0.6990 & 0.6823 & 0.8437 & 0.4149\\ 
\hline
% \bottomrule
\end{tabular}
}
\caption{Probability Distribution Function Non-overlap values between points and the five descriptors. Note that higher values imply higher disagreement between the two distributions. }
\label{table:2}
\end{table}

\noindent An examination of Table \ref{table:2} reveals that the greatest disagreement, over the time period considered, is between the profits and points achieved by a team. This is also clearly evident from Figure 4 where we re-plot the data against the seasons. Note that in Figure 4 the seasons run from right to left, with the data for 2009/10 displayed on the extreme right. We notice that the curve for the disagreement between the points and the profits is consistently above all others. 
(There was no data available for the profits and expenditure for the season 2017/18 when we undertook the analysis. Hence we disregard all entries for the season 2017/18 in Table \ref{table:2} above.)
\begin{figure}[H]
\center
	\centering
	\includegraphics[width=\textwidth]{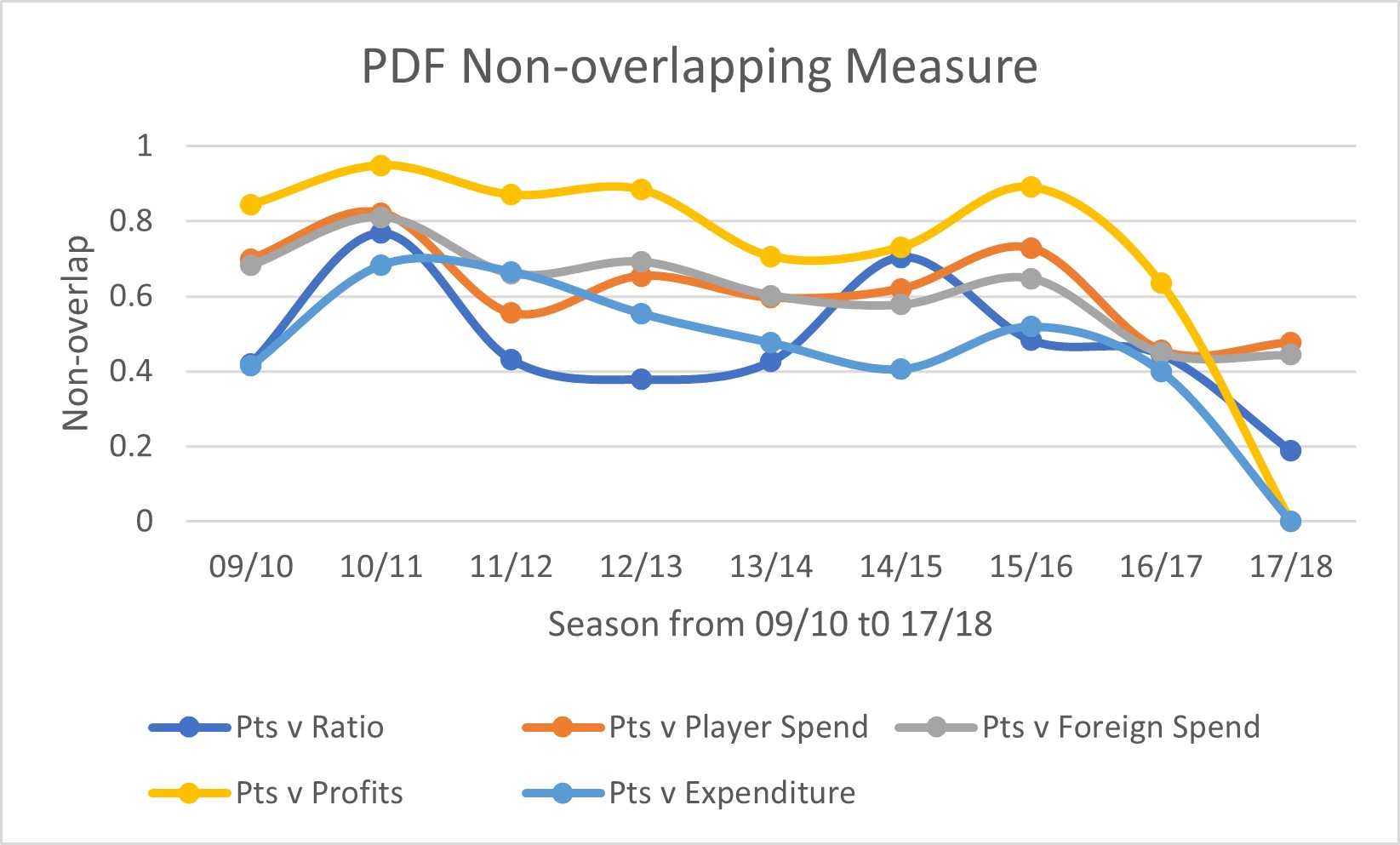}
	\centering
\caption{A plot of the non-overlapping measure of disagreement between points and the other descriptors spanning 2009/10 to 2017/18.}
\label{label:fig_nonoverlapping}
\end{figure}
\noindent Hence, the achievements of the teams on the playing fields are not translated into profits on the balance sheets for the clubs. In Figure \ref{label:fig_nonoverlapping}, we can also see that the peaks in all of the inequality curves occurred during the 2010/11 season. It's interesting to note that rules regarding financial fair play to encourage responsible spending for the long-term benefit of football were introduced in the 2011/12 season \cite{FairPlay}.
\subsection{Principle Component Analysis}
Finally, we perform a Principal Component Analysis (PCA) \cite{Gray} of the data in order to determine the most influential descriptors. Principal component analysis is a statistical procedure that uses an orthogonal transformation to convert a set of observations of correlated variables such as the ones we are dealing with into a set of values of linearly uncorrelated variables called principal components. The PCA was carried out using the standard inbuilt facility available in MATLAB 2023a \cite{Mat}. 

The outcomes from the PCA are tabulated in Appendix 2 (see Tables A9 - A16) for all of the seasons studied. The majority of the variation in the data is explained by the columns headed PCA 1. We can see that the expenditure has the highest coefficient in all of the cases tabulated, followed by expenditure on all players and foreign players. Hence, the majority of the variation between the teams can be described by these three descriptors. The results are consistent over all seasons considered.

\subsection{Impact of Descriptors on Economic Profit}
\begin{figure}[H]
\begin{subfigure}[b]{0.48\textwidth}
	\centering
	\includegraphics[width=\textwidth]{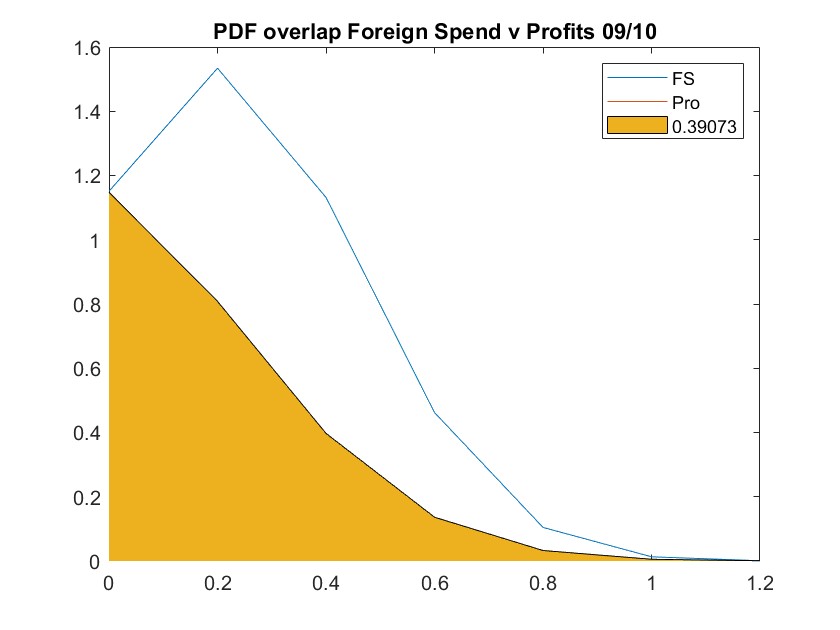}
	\caption{\centering Foreign spend to profit ratio = 39.07\%}
	\centering
	\label{label:file_name}
\end{subfigure}
\hfill
\begin{subfigure}[b]{0.48\textwidth}
	\centering
	\includegraphics[width=\textwidth]{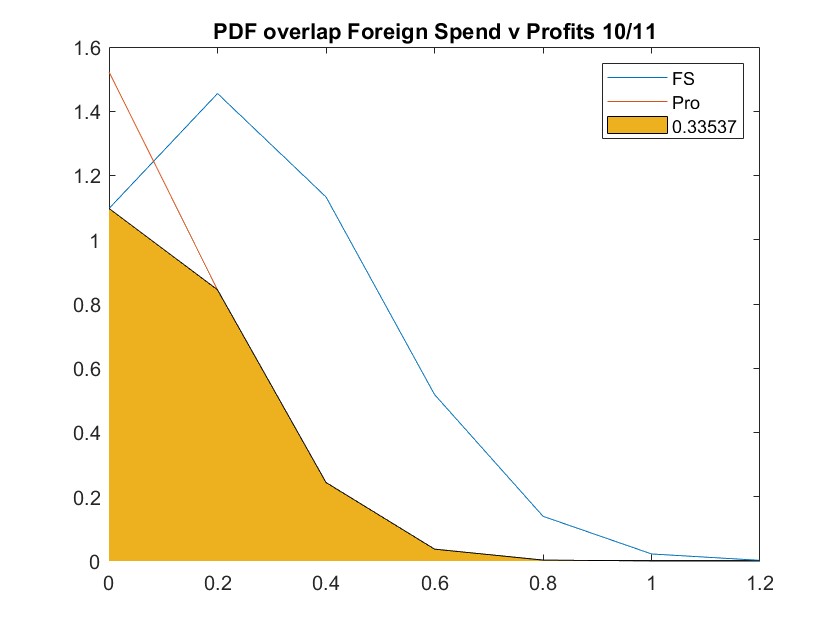}
	\caption{\centering Foreign spend to profit ratio = 33.54\%}
	\centering
	\label{label:file_name}
\end{subfigure}
\hfill
\begin{subfigure}[b]{0.48\textwidth}
	\centering
	\includegraphics[width=\textwidth]{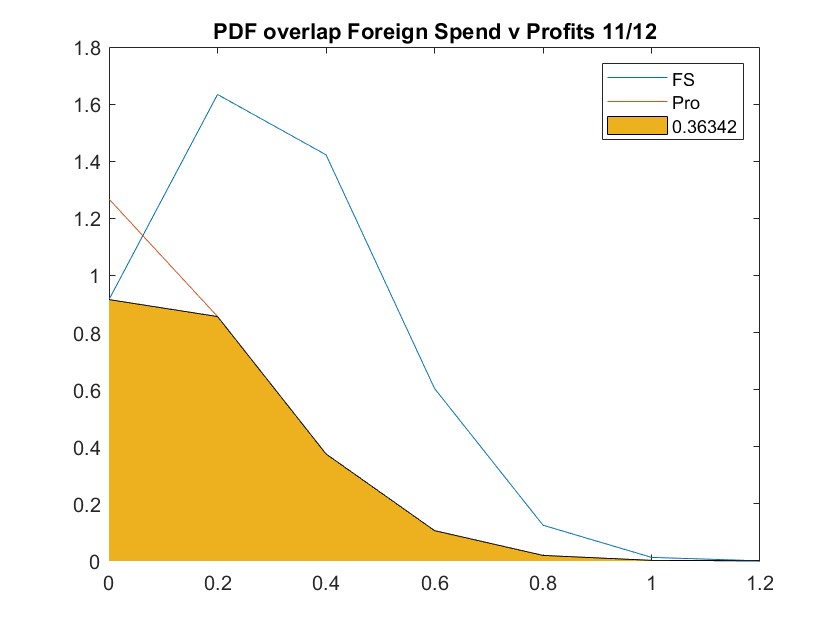}
	\caption{\centering Foreign spend to profit ratio = 36.34\%}
	\centering
	%\label{label:file_name}
\end{subfigure}
\hfill
\begin{subfigure}[b]{0.48\textwidth}
	\centering
	\includegraphics[width=\textwidth]{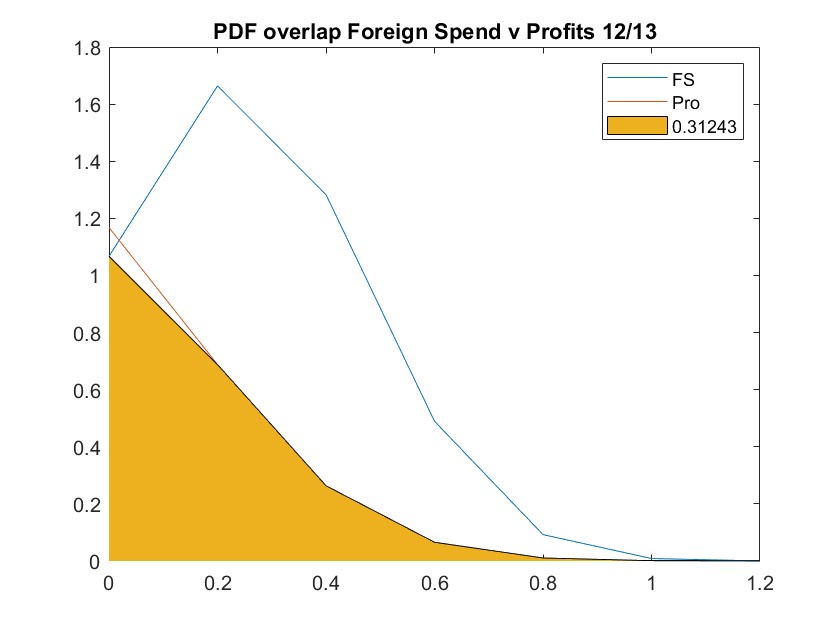}
	\caption{\centering Foreign spend to profit ratio = 31.24\%}
	\centering
	%\label{label:file_name}
	\end{subfigure}
	\hfill
	\begin{subfigure}[b]{0.48\textwidth}
	\centering
	\includegraphics[width=\textwidth]{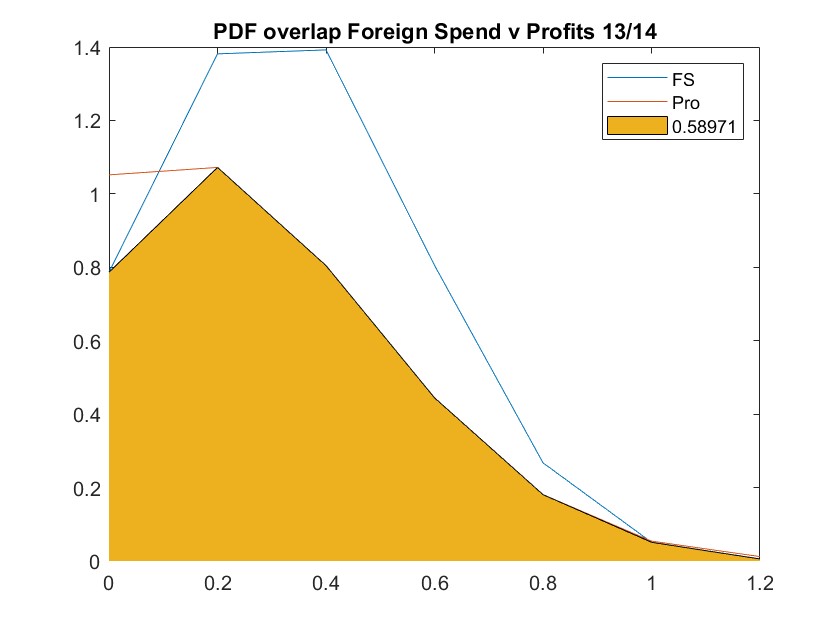}
	\caption{Foreign spend to profit ratio = 58.97\%}
	\centering
	\label{label:file_name}
\end{subfigure}
\hfill
\begin{subfigure}[b]{0.48\textwidth}
	\centering
	\includegraphics[width=\textwidth]{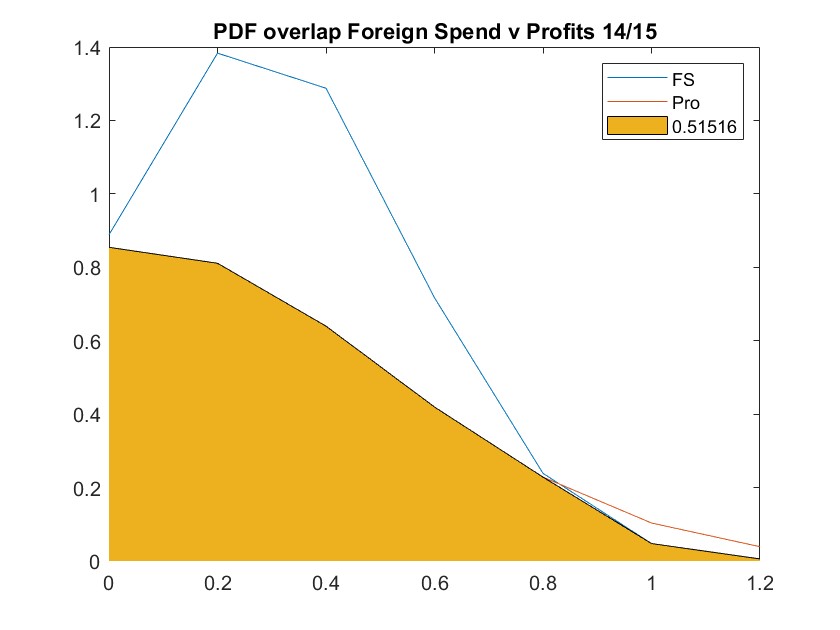}
	\caption{\centering Foreign spend to profit ratio = 51.52\%}
	\centering
	%\label{label:file_name}
\end{subfigure}
\hfill
\begin{subfigure}[b]{0.48\textwidth}
	\centering
	\includegraphics[width=\textwidth]{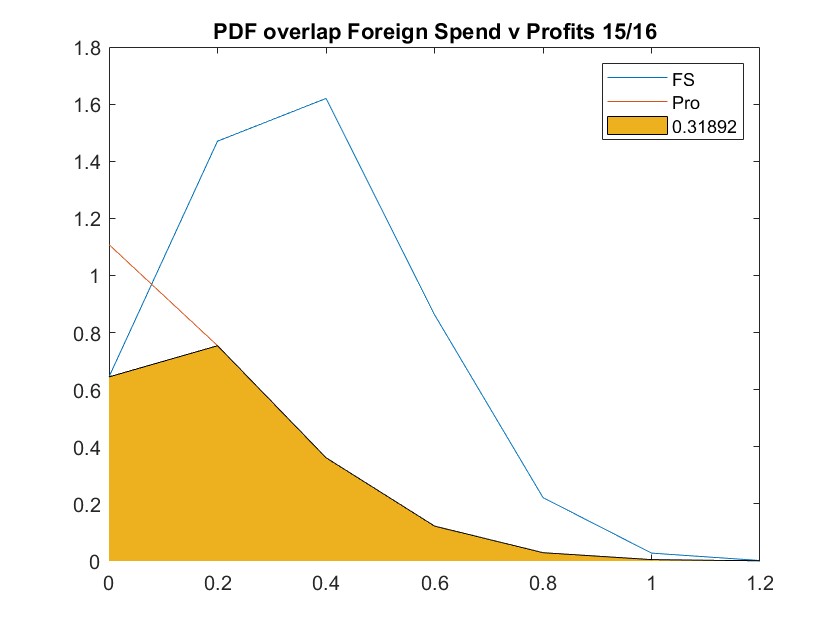}
	\caption{\centering Foreign spend to profit ratio = 31.89\%}
	\centering
	%\label{label:file_name}
\end{subfigure}
\hfill
\begin{subfigure}[b]{0.48\textwidth}
	\centering
	\includegraphics[width=\textwidth]{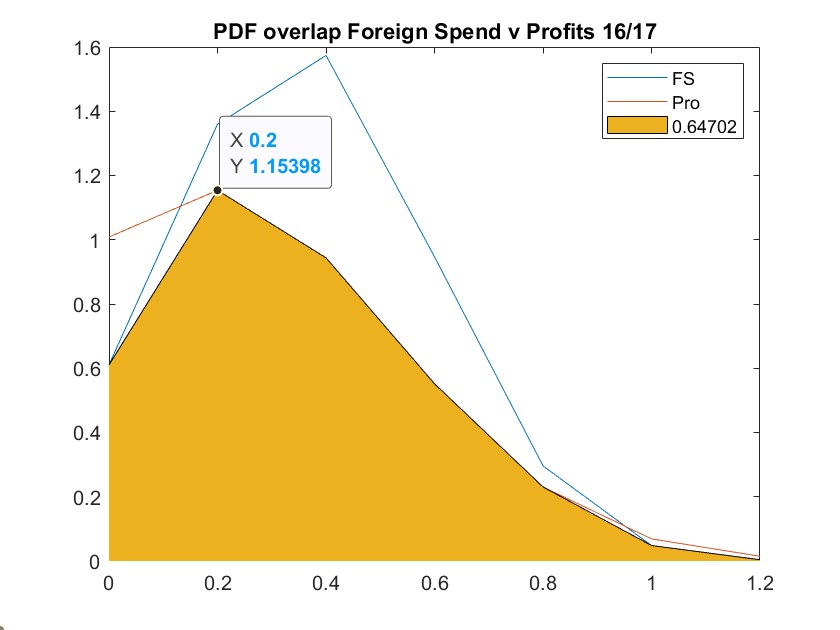}
	\caption{\centering Foreign spend to profit ratio = 64.70\%}
	\centering
	%\label{label:file_name}
\end{subfigure}
\caption{Percentage of spends on foreign players that translated into economic profit.}
\label{label:fig7}
\end{figure}

To analyze the quantitative impact on profit, the first of the indicators that we focused on evaluates the fraction of investment on foreign players that translated into eventual economic profit. To understand this, we calculate the overlapping regions of the probability density functions (PDFs) of the investment on foreign players to economic profits. Note, this is a complex question to answer as the 5 descriptors considered are interlinked and hence the impacts are difficult to unentangle. Nevertheless, as Figure \ref{label:fig7} clearly shows, 2015/16 was a remarkable year that contributed to a higher profit margin in the year following, compared to previous years. This is indicative of the impact of Leisceter City unexpectedly winning the Premier League against all odds in that season. 

\begin{figure}[H]
\begin{subfigure}[b]{0.48\textwidth}
	\centering
	\includegraphics[width=\textwidth]{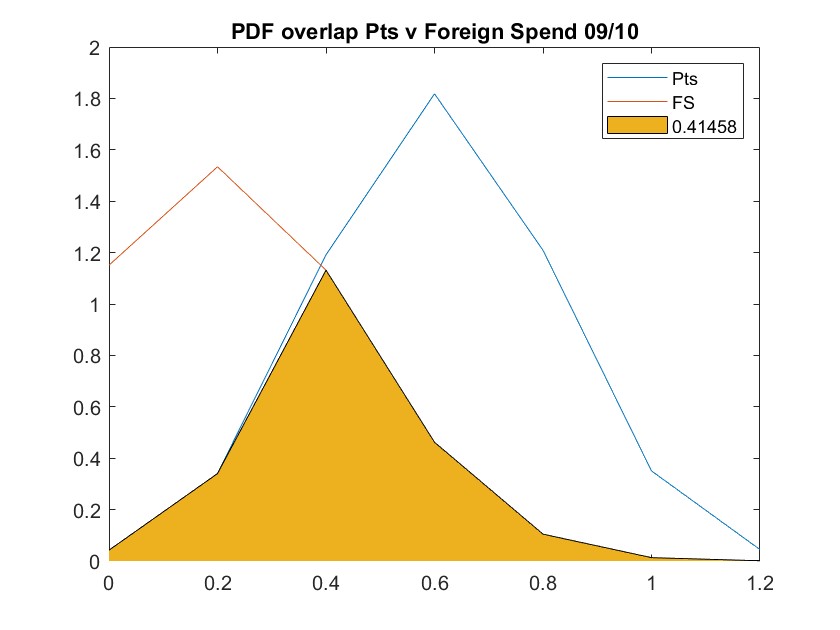}
	\caption{\centering Points to foreign spend ratio = 41.46\%}
	\centering
	\label{label:file_name}
\end{subfigure}
\hfill
\begin{subfigure}[b]{0.48\textwidth}
	\centering
	\includegraphics[width=\textwidth]{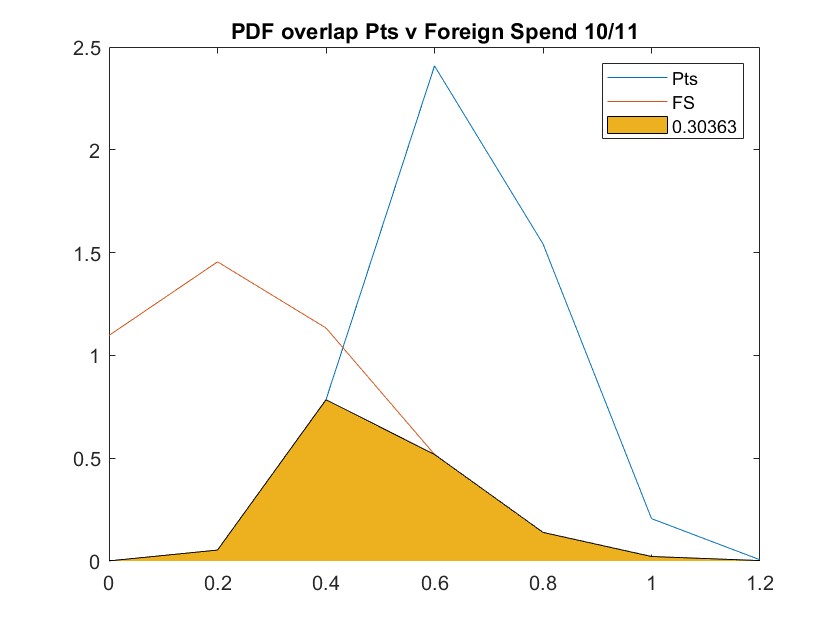}
	\caption{\centering Points to foreign spend ratio = 30.36\%}
	\centering
	%\label{label:file_name}
\end{subfigure}
\hfill
\begin{subfigure}[b]{0.48\textwidth}
	\centering
	\includegraphics[width=\textwidth]{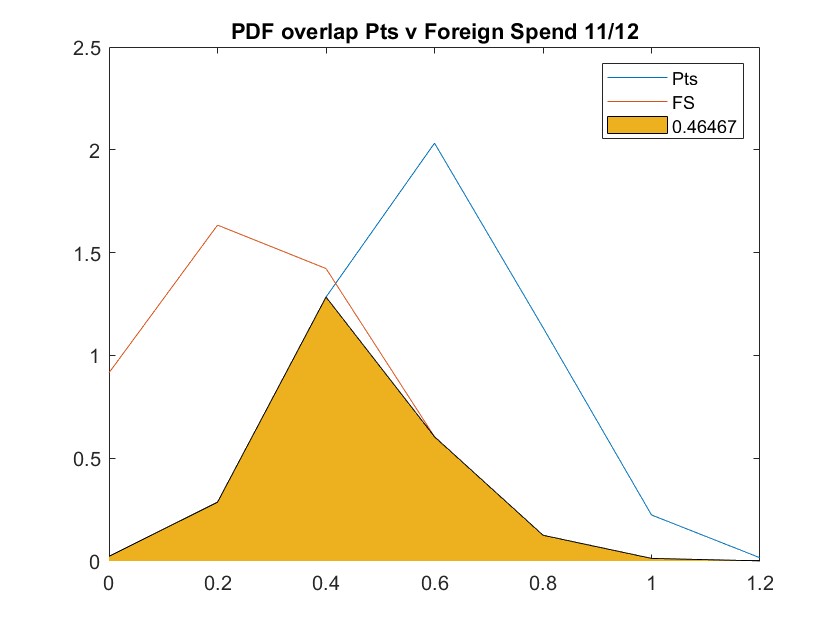}
	\caption{\centering Points to foreign spend ratio = 46.47\%}
	\centering
	%\label{label:file_name}
\end{subfigure}
\hfill
\begin{subfigure}[b]{0.48\textwidth}
	\centering
	\includegraphics[width=\textwidth]{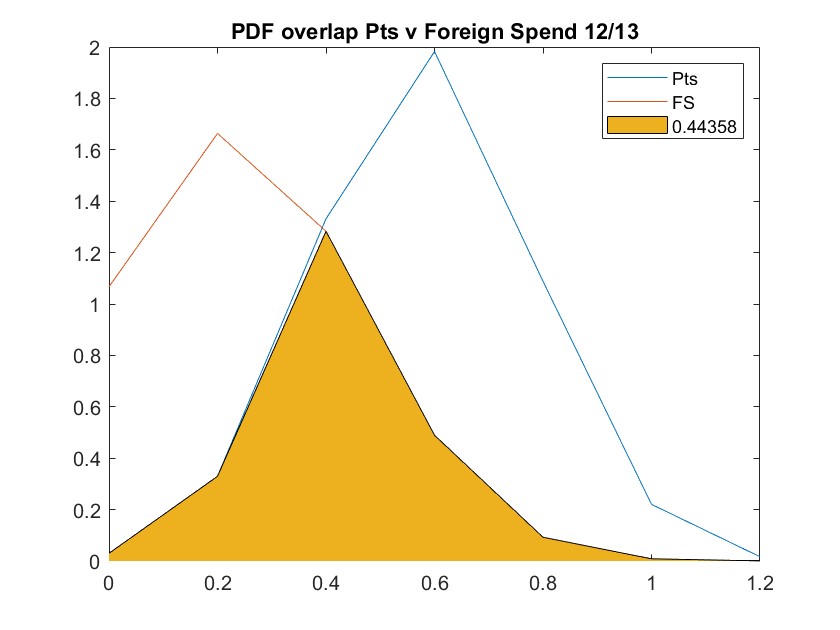}
	\caption{\centering Points to foreign spend ratio = 44.36\%}
	\centering
	%\label{label:file_name}
	\end{subfigure}
	\hfill
	\begin{subfigure}[b]{0.48\textwidth}
	\centering
	\includegraphics[width=\textwidth]{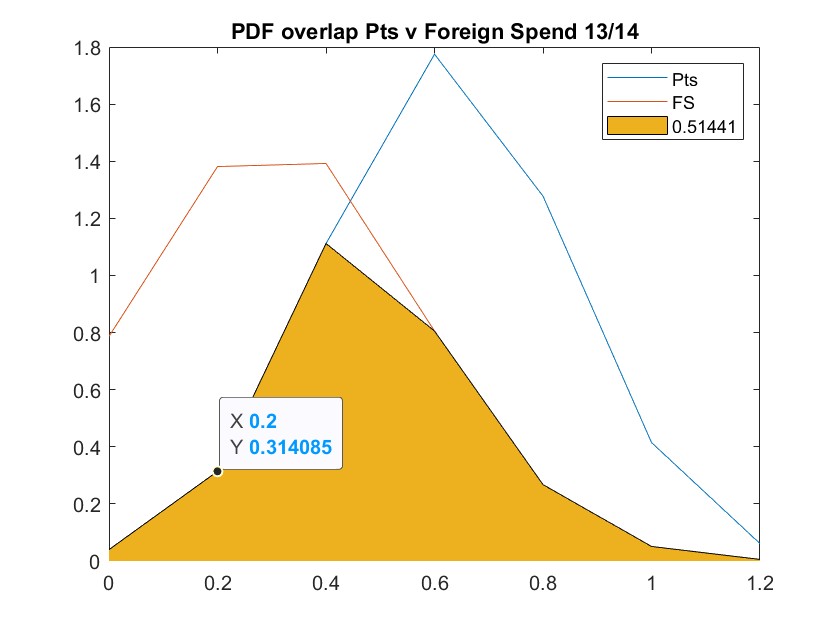}
	\caption{Points to foreign spend ratio = 51.44\%}
	\centering
	\label{label:file_name}
\end{subfigure}
\hfill
\begin{subfigure}[b]{0.48\textwidth}
	\centering
	\includegraphics[width=\textwidth]{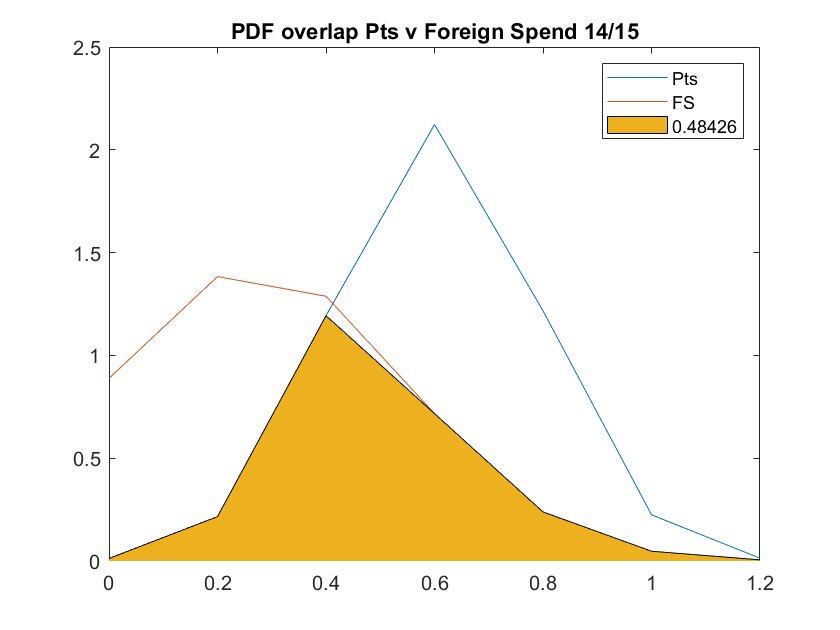}
	\caption{\centering Points to foreign spend ratio = 48.43\%}
	\centering
	%\label{label:file_name}
\end{subfigure}
\hfill
\begin{subfigure}[b]{0.48\textwidth}
	\centering
	\includegraphics[width=\textwidth]{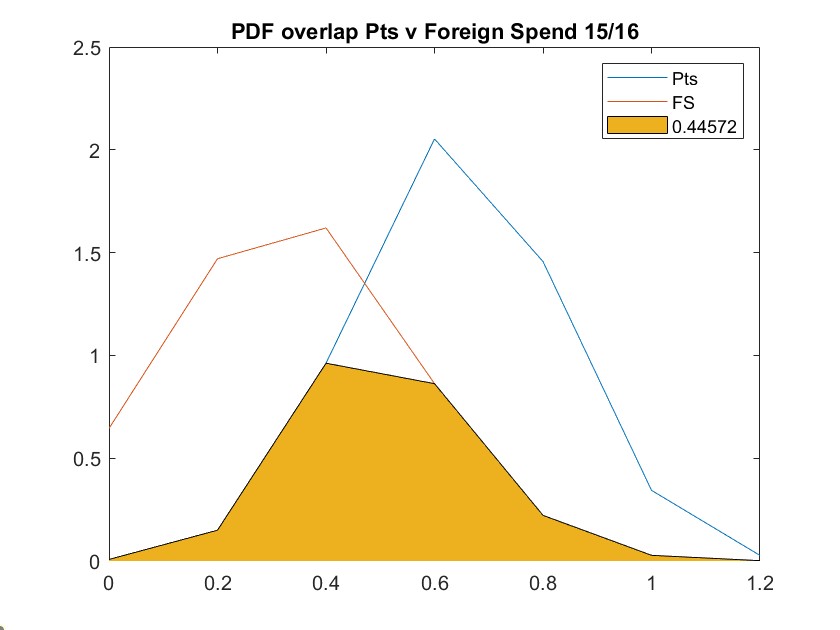}
	\caption{\centering Points to foreign spend ratio = 44.57\%}
	\centering
	%\label{label:file_name}
\end{subfigure}
\hfill
\begin{subfigure}[b]{0.48\textwidth}
	\centering
	\includegraphics[width=\textwidth]{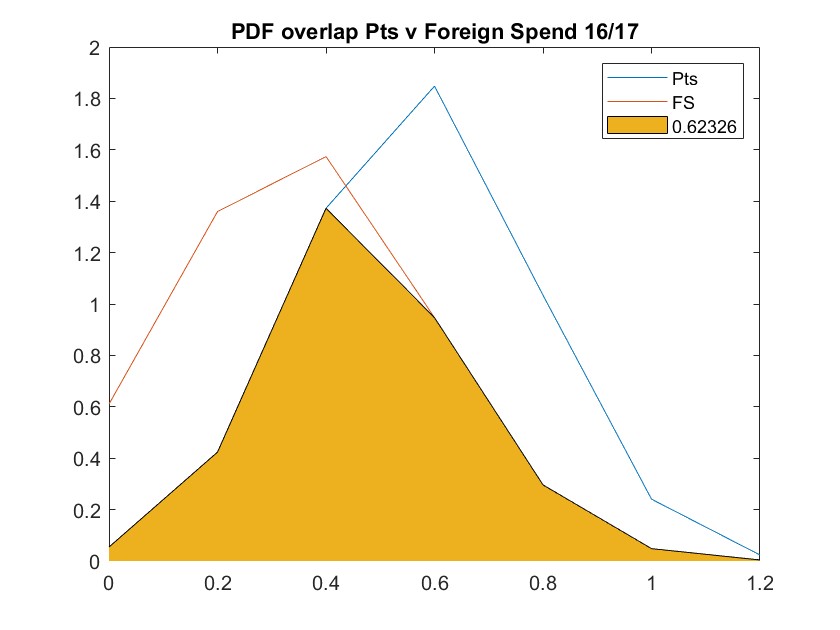}
	\caption{\centering Points to foreign spend ratio = 63.33\%}
	\centering
	%\label{label:file_name}
\end{subfigure}
%\hfill
%\begin{subfigure}[b]{0.48\textwidth}
%	\centering
%	\includegraphics[width=\textwidth]{PDF_overlap_points_vs_foreignspend_1718.jpg}
%	\caption{\centering Foreign spend to profit ratio = 60.04\%}
%	\centering
%	%\label{label:file_name}
%\end{subfigure}
\caption{Correlation between footballing performance, measured as number of points scrored, against economic profit.}
\label{label:fig8}
\end{figure}

A correlated indicator would be measuring the overlapping regimes of the probability density functions for the points scored against the investment on foreign players. This specific indicator is meant to understand how much of the investment on foreign players actually translate into performance on the field. As Figure \ref{label:fig8} shows, the correlation is clearly positive, an observation that follows the trend previously demonstrated in Figure \ref{label:fig7}. 

Three other indicators have also been measured. These are as follows: correlations between total investment on all players against profit, points acquired by the teams against investment on all players, and net economic profit against total expenditure. The trends are consistent with those from Figures \ref{label:fig7} and \ref{label:fig8}.

\section{Conclusion}
To conclude, we have undertaken an extensive analysis of the data for the English Premier League over the period covered from 2009/10 to 2016/17 (with some additional data from the 2017/18 season). Our analysis is structured on a multivariate assessment of five key descriptors, total expenditure, funds on all players, funds on foreign players, the ratio of the number of British to foreign players and overall monetary profit. The complexity of the mutual relationship between these factors is demonstrated in the mutual covariance match as shown in Figure \ref{label:fig2}. Clearly, there is no indicative pattern that can demonstrably comment on the impact trend. By re-ranking the teams using these five descriptors, we are able to establish alternative league tables to the conventional ones, ranking that could optimize investment against points procurement by the teams concerned. 

A comparative analysis of the differences between the tables permits us to establish the most likely factors to influence the performance of the clubs. We find that the top teams in the conventional league are also those that tend to have the highest expenditure overall as well as for all players and foreign players; they also have the highest ratios of foreign to British players, as clearly evidenced by the Gini coefficients calculated from the Lorenz curves shown in Figure \ref{label:fig_Gini} as also from the Theil indices as shown in Figure \ref{label:fig_Theil}. While the respective profiles for the ratio of British to foreign players and that for expenditure are closely matched over the years considered, the other three descriptors clearly show distinctive time dependence. Clearly, the impact from foreign players cannot be immediate. A certain minimal time would be required for investment on foreign players to mature into tangible productivity in the form of performance and profits. An open question here is a club-specific timeline for such productivity. However, we also find that a successful performance on the field by a team is not a guarantee for healthy profits at the end of the season. 

It is interesting to note that the season of 2015/16, when Leicester City triumphed, is an exception to our overall conclusions. Indeed, it could be argued that 2015/16 has been the most successful season for British football over the period considered as the correlations for the ratio of foreign to British players, the spend on foreign players, the total spend and the total expenditure were all at their lowest. Furthermore, the correlation for the profits was positive. All such details can be found in the 3 Appendices.

Our analysis indicates that on-field performances do not necessarily reflect as economic profits or losses. It would be interesting to investigate whether our findings are replicated in other football leagues around the world. We are also planning an extension of this analysis to incorporate additional strategic factors at granular levels, that would be club-specific and targeted towards optimizing footballing and economic performance focusing on sustainability while still being competitive. Furthermore, our study can in principle be used by the football club owners to target either glory on the field or a healthy balance sheet.

\newpage
\appendix 
\addcontentsline{toc}{chapter}{Appendix}
\section*{\large Appendix 1: Re-ranked Tables For All Twenty Teams Over Eight Seasons (2017/18 data are only partially available)}
\setcounter{table}{0}
\renewcommand{\thetable}{A\arabic{table}}
\renewcommand{\thefigure}{F\arabic{figure}}
% \textbf{}

\begin{table}[ht!]
\centering
\resizebox{\linewidth}{!}{%
\begin{tabular}{|c|c|c|c|c|c|c|c|c|c|c|c|c|}
\hline
 Expense & Profit & Total & Foreign & Foreign: & Position & Team & Pts & Ratio & Total  & Foreign & Profits & Exp\\ 
 (\pounds m) & (\pounds m) & spend (\pounds m) & spend (\pounds m) & British  &  & &  &  & spend (\pounds m) & spend (\pounds m) & (\pounds m) & (\pounds m) \\
 %\midrule
\hline 
18 & 9 & 18 & 19 & 20 & 1 & Chelsea & 93 & 3.8 & 143 & 143 & 12.97 & 389.927 \\ 
15 & 13 & 15 & 13 & 10 & 2 & Tottenham & 86 & 1.3077 & 91.9 & 80.9 & 28.063 & 283.953 \\ 
20 & 4 & 20 & 18 & 19 & 3 & Manchester City & 78 & 3.6667 & 194.3 & 138.8 & 1.088 & 512.375 \\ 
17 & 17 & 12 & 14 & 6 & 4 & Liverpool & 76 & 1 & 82.1 & 82.1 & 38.917 & 365.246 \\ 
16 & 20 & 14 & 16 & 15 & 5 & Arsenal & 75 & 1.75 & 88.4 & 86 & 58.897 & 345.305 \\ 
19 & 1 & 19 & 20 & 8 & 6 & Manchester United & 69 & 1.0625 & 173.3 & 173.3 & -85.701 & 414.172 \\ 
14 & 14 & 13 & 12 & 9 & 7 & Everton & 61 & 1.0714 & 84.15 & 70.15 & 30.617 & 194.466 \\ 
13 & 16 & 11 & 11 & 5 & 8 & Southampton & 46 & 0.83333 & 65.5 & 52 & 34.632 & 187.471 \\ 
5 & 11 & 5 & 5 & 3 & 9 & Bournemouth & 46 & 0.375 & 35.1 & 28.1 & 13.991 & 125.028 \\ 
4 & 15 & 2 & 2 & 1 & 10 & West Bromwich & 45 & 0.35294 & 21.7 & 15.2 & 32.289 & 119.751 \\ 
11 & 19 & 10 & 10 & 14 & 11 & West Ham & 45 & 1.6154 & 52.2 & 40.2 & 43.041 & 170.895 \\ 
1 & 3 & 17 & 17 & 16 & 12 & Leicester & 44 & 2 & 97 & 97 & -0.593 & 20.928 \\ 
7 & 5 & 6 & 7 & 17 & 13 & Stoke & 44 & 2.1111 & 35.9 & 34.4 & 3.574 & 136.169 \\ 
12 & 7 & 16 & 15 & 7 & 14 & Crystal Palace & 41 & 1 & 96.9 & 84.9 & 10.073 & 171.417 \\ 
9 & 10 & 7 & 6 & 12 & 15 & Swansea & 41 & 1.3846 & 39.2 & 34 & 12.996 & 152.291 \\ 
2 & 12 & 4 & 3 & 2 & 16 & Burnley & 40 & 0.36842 & 29.9 & 24.1 & 22.134 & 101.025 \\ 
8 & 6 & 8 & 9 & 18 & 17 & Watford & 40 & 2.2 & 39.25 & 39.25 & 8.141 & 142.322 \\ 
3 & 18 & 1 & 1 & 4 & 18 & Hull & 34 & 0.77778 & 9 & 9 & 42.46 & 112.593 \\ 
6 & 8 & 9 & 8 & 13 & 19 & Middlesbrough & 28 & 1.5 & 43.55 & 37.2 & 11.486 & 125.906 \\ 
10 & 2 & 3 & 4 & 11 & 20 & Sunderland & 24 & 1.3333 & 25.5 & 25.5 & -10.248 & 170.616 \\ 
\hline 
\end{tabular}
}
\caption{Re-rank table for 2016/17 Season. Total Player Spend, Total Foreign Spend, Profits and Expenditure are in millions.}
\label{table:A2}
\end{table}

\begin{table}[ht!]
\centering
\resizebox{\linewidth}{!}{%
\begin{tabular}{|c|c|c|c|c|c|c|c|c|c|c|c|c|}
\hline
 Expense & Profit & Total & Foreign & Foreign: & Position & Team & Pts & Ratio & Total  & Foreign & Profits & Exp\\ 
 (\pounds m) & (\pounds m) & spend (\pounds m) & spend (\pounds m) & British  &  & &  &  & spend (\pounds m) & spend (\pounds m) & (\pounds m) & (\pounds m) \\
 %\midrule
\hline 
1 & 10 & 7 & 7 & 5 & 1 & Leicester & 81 & 0.8125 & 34.6 & 29.6 & -0.828 & 19.848 \\ 
16 & 20 & 3 & 4 & 17 & 2 & Arsenal & 71 & 2.2 & 20.6 & 20.6 & 25.496 & 315.946 \\ 
15 & 18 & 16 & 15 & 11 & 3 & Tottenham & 70 & 1 & 69.7 & 64.8 & 15.051 & 207.843 \\ 
19 & 19 & 20 & 19 & 19 & 4 & Manchester City & 66 & 2.7778 & 201.9 & 132.5 & 20.483 & 395.336 \\ 
17 & 1 & 19 & 20 & 10 & 5 & Manchester United & 66 & 0.95 & 145 & 145 & -82.72 & 344.393 \\ 
12 & 16 & 10 & 10 & 14 & 6 & Southampton & 63 & 1.3333 & 43.75 & 38.5 & 4.972 & 152.514 \\ 
13 & 9 & 12 & 13 & 9 & 7 & West Ham & 62 & 0.94737 & 49.3 & 44.5 & -4.876 & 152.85 \\ 
18 & 6 & 18 & 18 & 8 & 8 & Liverpool & 60 & 0.91304 & 110 & 92.4 & -21.391 & 366.914 \\ 
6 & 12 & 5 & 6 & 20 & 9 & Stoke & 51 & 2.8571 & 25 & 25 & 2.055 & 116.566 \\ 
20 & 2 & 15 & 16 & 18 & 10 & Chelsea & 50 & 2.75 & 69.65 & 69.65 & -72.347 & 423.13 \\ 
14 & 5 & 8 & 8 & 6 & 11 & Everton & 47 & 0.83333 & 37.3 & 37.3 & -24.333 & 155.043 \\ 
7 & 7 & 1 & 2 & 7 & 12 & Swansea & 47 & 0.88235 & 10.5 & 10.5 & -13.081 & 118.407 \\ 
3 & 14 & 4 & 5 & 16 & 13 & Watford & 45 & 1.7692 & 22.5 & 22.5 & 3.613 & 96.472 \\ 
4 & 11 & 9 & 11 & 1 & 14 & West Bromwich & 43 & 0.57895 & 42.95 & 38.65 & 0.999 & 100.296 \\ 
8 & 8 & 6 & 3 & 4 & 15 & Crystal Palace & 42 & 0.8 & 26.8 & 14 & -5.567 & 118.425 \\ 
2 & 13 & 13 & 9 & 2 & 16 & Bournemouth & 42 & 0.6087 & 49.71 & 38.41 & 3.395 & 95.988 \\ 
11 & 3 & 11 & 12 & 13 & 17 & Sunderland & 39 & 1.2857 & 46.6 & 43.8 & -33.785 & 142.913 \\ 
9 & 15 & 17 & 17 & 15 & 18 & Newcastle & 37 & 1.5385 & 87.7 & 71.7 & 4.597 & 124.852 \\ 
5 & 17 & 2 & 1 & 3 & 19 & Norwich & 34 & 0.63158 & 13.2 & 3 & 9.415 & 112.983 \\ 
10 & 4 & 14 & 14 & 12 & 20 & Aston Villa & 17 & 1.0588 & 61.3 & 57.9 & -29.654 & 137.709 \\ 
\hline 
\end{tabular}
}
\caption{Re-rank table for 2015/16 Season. Total Player Spend, Total Foreign Spend, Profits and Expenditure are in millions.}
\label{table:A3}
\end{table}

\begin{table}[ht!]
\centering
\resizebox{\linewidth}{!}{%
\begin{tabular}{|c|c|c|c|c|c|c|c|c|c|c|c|c|}
\hline
 Expense & Profit & Total & Foreign & Foreign: & Position & Team & Pts & Ratio & Total  & Foreign & Profits & Exp\\ 
 (\pounds m) & (\pounds m) & spend (\pounds m) & spend (\pounds m) & British  &  & &  &  & spend (\pounds m) & spend (\pounds m) & (\pounds m) & (\pounds m) \\
 %\midrule 
\hline 
19 & 4 & 19 & 19 & 19 & 1 & Chelsea & 87 & 2.5 & 140.3 & 140.3 & -26.279 & 353 \\ 
20 & 14 & 17 & 17 & 20 & 2 & Manchester City & 79 & 4.4 & 113 & 113 & 10.54 & 355.781 \\ 
18 & 19 & 16 & 16 & 15 & 3 & Arsenal & 75 & 1.6429 & 107.7 & 87.7 & 51.833 & 305.871 \\ 
17 & 1 & 20 & 20 & 14 & 4 & Manchester United & 70 & 1.2308 & 182.5 & 148.5 & -65.712 & 295.78 \\ 
15 & 10 & 10 & 9 & 16 & 5 & Tottenham & 64 & 1.6667 & 23.8 & 18.8 & 2.544 & 198.627 \\ 
16 & 20 & 18 & 18 & 13 & 6 & Liverpool & 62 & 1.0588 & 133.3 & 128.3 & 58.762 & 295.223 \\ 
13 & 15 & 15 & 13 & 8 & 7 & Southampton & 60 & 0.875 & 49 & 36.5 & 12.257 & 143.808 \\ 
9 & 7 & 6 & 8 & 3 & 8 & Swansea & 56 & 0.7 & 17.75 & 15.5 & 0.685 & 121.746 \\ 
4 & 12 & 2 & 3 & 18 & 9 & Stoke & 54 & 2.25 & 3.8 & 3.8 & 5.246 & 96.116 \\ 
5 & 13 & 5 & 4 & 4 & 10 & Crystal Palace & 48 & 0.7 & 13 & 4 & 6.352 & 98.512 \\ 
12 & 5 & 13 & 15 & 9 & 11 & Everton & 47 & 0.88889 & 40.3 & 40.3 & -4.607 & 133.507 \\ 
10 & 11 & 11 & 11 & 10 & 12 & West Ham & 47 & 1 & 24.9 & 22.4 & 4.053 & 122.68 \\ 
6 & 9 & 7 & 7 & 2 & 13 & West Bromwich & 44 & 0.625 & 19 & 12.6 & 1.344 & 100.334 \\ 
1 & 8 & 9 & 10 & 6 & 14 & Leicester & 41 & 0.76471 & 21.7 & 21.7 & 0.982 & 22.529 \\ 
8 & 18 & 12 & 12 & 17 & 15 & Newcastle & 39 & 1.6667 & 27.2 & 27.2 & 32.528 & 114.295 \\ 
11 & 3 & 4 & 2 & 5 & 16 & Sunderland & 38 & 0.72222 & 12.6 & 0 & -26.677 & 127.559 \\ 
14 & 2 & 3 & 6 & 11 & 17 & Aston Villa & 38 & 1 & 12.5 & 12.5 & -57.346 & 169.796 \\ 
3 & 16 & 14 & 14 & 12 & 18 & Hull & 35 & 1 & 45 & 37 & 20.395 & 88.129 \\ 
2 & 17 & 1 & 1 & 1 & 19 & Burnley & 33 & 0.3 & 3.8 & 0 & 29.932 & 49.508 \\ 
7 & 6 & 8 & 5 & 7 & 20 & QPR & 30 & 0.84211 & 20 & 10 & 0.467 & 113.549 \\ 
\hline 
\end{tabular}
}
\caption{Re-rank table for 2014/15 Season. Total Player Spend, Total Foreign Spend, Profits and Expenditure are in millions.}
\label{table:A4}
\end{table}

\begin{table}[ht!]
\centering
\resizebox{\linewidth}{!}{%
\begin{tabular}{|c|c|c|c|c|c|c|c|c|c|c|c|c|}
\hline
 Expense & Profit & Total & Foreign & Foreign: & Position & Team & Pts & Ratio & Total  & Foreign & Profits & Exp\\ 
 (\pounds m) & (\pounds m) & spend (\pounds m) & spend (\pounds m) & British  &  & &  &  & spend (\pounds m) & spend (\pounds m) & (\pounds m) & (\pounds m) \\
 %\midrule
\hline 
20 & 3 & 18 & 18 & 16 & 1 & Manchester City & 86 & 1.9 & 80.6 & 80.6 & -22.929 & 371.465 \\ 
16 & 6 & 14 & 14 & 12 & 2 & Liverpool & 84 & 1.3846 & 38.5 & 38.5 & 0.413 & 255.25 \\ 
19 & 14 & 19 & 19 & 19 & 3 & Chelsea & 82 & 3 & 82.7 & 82.7 & 14.301 & 342.276 \\ 
17 & 19 & 16 & 16 & 20 & 4 & Arsenal & 79 & 3 & 50 & 50 & 48.397 & 261.465 \\ 
11 & 17 & 10 & 13 & 7 & 5 & Everton & 72 & 1 & 22.8 & 22.8 & 28.232 & 120.49 \\ 
15 & 20 & 20 & 20 & 10 & 6 & Tottenham & 69 & 1.0556 & 111.6 & 111.6 & 69.666 & 201.008 \\ 
18 & 2 & 17 & 17 & 8 & 7 & Manchester United & 64 & 1 & 79.5 & 79.5 & -32.033 & 280.124 \\ 
10 & 18 & 15 & 15 & 3 & 8 & Southampton & 56 & 0.70588 & 42.5 & 42.5 & 30.973 & 111.296 \\ 
5 & 9 & 1 & 1 & 17 & 9 & Stoke & 50 & 2 & 3.5 & 3.5 & 3.786 & 94.557 \\ 
13 & 16 & 4 & 4 & 15 & 10 & Newcastle & 49 & 1.7273 & 13 & 13 & 18.752 & 125.008 \\ 
1 & 15 & 3 & 3 & 1 & 11 & Crystal Palace & 45 & 0.45455 & 11.9 & 8.25 & 17.939 & 72.652 \\ 
7 & 8 & 11 & 6 & 6 & 12 & Swansea & 42 & 0.95 & 23.3 & 17.4 & 1.733 & 97.431 \\ 
8 & 12 & 12 & 9 & 4 & 13 & West Ham & 40 & 0.71429 & 25 & 18 & 10.345 & 107.733 \\ 
12 & 4 & 8 & 7 & 13 & 14 & Sunderland & 38 & 1.3846 & 21.45 & 17.8 & -16.313 & 122.137 \\ 
9 & 7 & 6 & 10 & 14 & 15 & Aston Villa & 38 & 1.4615 & 19.3 & 19.3 & 1.577 & 108.991 \\ 
2 & 11 & 5 & 5 & 11 & 16 & Hull & 37 & 1.1429 & 17.6 & 15.6 & 9.409 & 81.409 \\ 
3 & 13 & 2 & 2 & 9 & 17 & West Bromwich & 36 & 1 & 6 & 6 & 10.761 & 85.705 \\ 
4 & 10 & 9 & 11 & 5 & 18 & Norwich & 33 & 0.82353 & 21.8 & 19.3 & 6.74 & 90.576 \\ 
14 & 1 & 7 & 12 & 18 & 19 & Fulham & 32 & 2 & 21.1 & 21.1 & -33.306 & 125.701 \\ 
6 & 5 & 13 & 8 & 2 & 20 & Cardiff & 30 & 0.61905 & 27.9 & 17.9 & -11.718 & 96.035 \\ 
\hline 
\end{tabular}
}
\caption{Re-rank table for 2013/14 Season. Total Player Spend, Total Foreign Spend, Profits and Expenditure are in millions.}
\label{table:A5}
\end{table}

\begin{table}[ht]
\centering
\resizebox{\linewidth}{!}{%
\begin{tabular}{|c|c|c|c|c|c|c|c|c|c|c|c|c|}
\hline
 Expense & Profit & Total & Foreign & Foreign: & Position & Team & Pts & Ratio & Total  & Foreign & Profits & Exp\\ 
 (\pounds m) & (\pounds m) & spend (\pounds m) & spend (\pounds m) & British  &  & &  &  & spend (\pounds m) & spend (\pounds m) & (\pounds m) & (\pounds m) \\
 %\midrule
\hline 
16 & 7 & 18 & 18 & 6 & 1 & Manchester United & 89 & 0.82353 & 53.75 & 50 & -11.372 & 227.483 \\ 
20 & 1 & 11 & 12 & 16 & 2 & Manchester City & 78 & 2 & 26.6 & 18.8 & -79.368 & 404.201 \\ 
19 & 2 & 20 & 20 & 19 & 3 & Chelsea & 75 & 2.5 & 108.9 & 108.9 & -56.068 & 298.975 \\ 
17 & 19 & 19 & 19 & 18 & 4 & Arsenal & 73 & 2.3636 & 58 & 58 & 13.216 & 251.812 \\ 
15 & 8 & 16 & 17 & 7 & 5 & Tottenham & 72 & 0.82353 & 35.8 & 35.8 & -8.938 & 167.736 \\ 
13 & 16 & 9 & 9 & 13 & 6 & Everton & 63 & 1.2308 & 18.35 & 14.85 & 1.597 & 100.453 \\ 
18 & 3 & 17 & 15 & 11 & 7 & Liverpool & 61 & 0.94444 & 46 & 27 & -46.85 & 255.979 \\ 
3 & 17 & 4 & 1 & 10 & 8 & West Bromwich & 49 & 0.86667 & 5 & 0 & 5.333 & 67.54 \\ 
4 & 20 & 3 & 3 & 4 & 9 & Swansea & 46 & 0.68421 & 3.8 & 2.5 & 15.277 & 73.326 \\ 
7 & 10 & 10 & 7 & 5 & 10 & West Ham & 46 & 0.78947 & 22.7 & 9.2 & -3.511 & 94.536 \\ 
6 & 14 & 7 & 8 & 2 & 11 & Norwich & 44 & 0.47059 & 11.9 & 11.9 & 0.523 & 80.089 \\ 
12 & 11 & 5 & 6 & 20 & 12 & Fulham & 43 & 2.875 & 8.5 & 6 & -2.71 & 98.509 \\ 
11 & 4 & 6 & 5 & 8 & 13 & Stoke & 42 & 0.85714 & 9.9 & 4.9 & -31.119 & 97.826 \\ 
5 & 9 & 15 & 13 & 3 & 14 & Southampton & 41 & 0.55 & 35.1 & 26.3 & -5.303 & 75.549 \\ 
9 & 5 & 8 & 10 & 14 & 15 & Aston Villa & 41 & 1.6667 & 16.13 & 15.5 & -15.331 & 97.016 \\ 
10 & 18 & 13 & 16 & 15 & 16 & Newcastle & 41 & 1.6923 & 29.8 & 29.8 & 8.814 & 97.64 \\ 
8 & 6 & 14 & 11 & 1 & 17 & Sunderland & 39 & 0.43478 & 33.3 & 18.3 & -13.144 & 96.397 \\ 
1 & 13 & 2 & 4 & 17 & 18 & Wigan & 36 & 2 & 3.5 & 3.5 & -0.684 & 4.03 \\ 
2 & 12 & 1 & 2 & 9 & 19 & Reading & 28 & 0.85714 & 1.11 & 0.74 & -2.339 & 63.077 \\ 
14 & 15 & 12 & 14 & 12 & 20 & QPR & 25 & 1.2 & 26.7 & 26.7 & 0.616 & 104.849 \\ 
\hline 
\end{tabular}
}
\caption{Re-rank table for 2012/13 Season. Total Player Spend, Total Foreign Spend, Profits and Expenditure are in millions.}
\label{table:A6}
\end{table}

\begin{table}[ht!]
\centering
\resizebox{\linewidth}{!}{%
\begin{tabular}{|c|c|c|c|c|c|c|c|c|c|c|c|c|}
\hline
 Expense & Profit & Total & Foreign & Foreign: & Position & Team & Pts & Ratio & Total  & Foreign & Profits & Exp\\ 
 (\pounds m) & (\pounds m) & spend (\pounds m) & spend (\pounds m) & British  &  & &  &  & spend (\pounds m) & spend (\pounds m) & (\pounds m) & (\pounds m) \\
 %\midrule
\hline 
20 & 1 & 20 & 20 & 20 & 1 & Manchester City & 89 & 3 & 78.8 & 78.8 & -95.296 & 352.546 \\ 
16 & 10 & 17 & 17 & 5 & 2 & Manchester United & 89 & 0.72222 & 55.7 & 37 & -2.278 & 206.774 \\ 
18 & 20 & 16 & 15 & 19 & 3 & Arsenal & 70 & 2.875 & 44.4 & 27.1 & 59.509 & 230.758 \\ 
15 & 9 & 2 & 1 & 3 & 4 & Tottenham & 69 & 0.71429 & 5.7 & 0 & -4.805 & 151.032 \\ 
10 & 15 & 13 & 14 & 13 & 5 & Newcastle & 65 & 1.3077 & 25.7 & 25.7 & 1.401 & 98.39 \\ 
19 & 13 & 19 & 19 & 16 & 6 & Chelsea & 64 & 1.8 & 60.5 & 52 & 0.2 & 277.201 \\ 
13 & 8 & 6 & 9 & 17 & 7 & Everton & 56 & 1.8 & 6.6 & 6.6 & -9.106 & 103.89 \\ 
17 & 2 & 18 & 18 & 6 & 8 & Liverpool & 52 & 0.76471 & 59.7 & 37.2 & -40.522 & 209.579 \\ 
12 & 6 & 11 & 12 & 18 & 9 & Fulham & 52 & 2.6667 & 15.6 & 15.6 & -18.755 & 102.201 \\ 
6 & 11 & 7 & 6 & 10 & 10 & West Bromwich & 47 & 0.9375 & 8.1 & 5.1 & -0.367 & 73.08 \\ 
2 & 19 & 8 & 8 & 2 & 11 & Swansea & 47 & 0.47826 & 8.58 & 6.28 & 14.632 & 50.735 \\ 
4 & 18 & 3 & 3 & 1 & 12 & Norwich & 47 & 0.30435 & 5.8 & 2.2 & 13.467 & 62.599 \\ 
14 & 3 & 12 & 10 & 9 & 13 & Sunderland & 45 & 0.8125 & 15.9 & 9.3 & -31.013 & 108.06 \\ 
8 & 7 & 14 & 13 & 8 & 14 & Stoke & 45 & 0.8 & 27.1 & 15.8 & -9.529 & 82.363 \\ 
1 & 12 & 5 & 7 & 15 & 15 & Wigan & 43 & 1.7 & 6.6 & 5.5 & -0.24 & 3.603 \\ 
11 & 5 & 15 & 16 & 11 & 16 & Aston Villa & 38 & 0.9375 & 29.8 & 29.8 & -19.944 & 99.7 \\ 
7 & 14 & 10 & 11 & 12 & 17 & QPR & 37 & 1 & 12.8 & 9.8 & 0.273 & 74.629 \\ 
9 & 4 & 4 & 5 & 4 & 18 & Bolton & 36 & 0.71429 & 6.6 & 3 & -22.117 & 91.274 \\ 
5 & 16 & 9 & 4 & 14 & 19 & Blackburn & 31 & 1.4286 & 11 & 2.3 & 4.289 & 72.832 \\ 
3 & 17 & 1 & 2 & 7 & 20 & Wolverhampton & 25 & 0.78947 & 5.6 & 0 & 5.76 & 56.979 \\ 
\hline 
\end{tabular}
}
\caption{Re-rank table for 2011/12 Season. Total Player Spend, Total Foreign Spend, Profits and Expenditure are in millions.}
\label{table:A7}
\end{table}

\begin{table}[ht]
\centering
\resizebox{\linewidth}{!}{%
\begin{tabular}{|c|c|c|c|c|c|c|c|c|c|c|c|c|}
\hline
 Expense & Profit & Total & Foreign & Foreign: & Position & Team & Pts & Ratio & Total  & Foreign & Profits & Exp\\ 
 (\pounds m) & (\pounds m) & spend (\pounds m) & spend (\pounds m) & British  &  & &  &  & spend (\pounds m) & spend (\pounds m) & (\pounds m) & (\pounds m) \\
 %\midrule
\hline 
17 & 16 & 11 & 11 & 11 & 1 & Manchester United & 80 & 1.1765 & 9 & 9 & 9.457 & 208.172 \\ 
19 & 2 & 19 & 19 & 18 & 2 & Chelsea & 71 & 2.375 & 110 & 110 & -71.962 & 295.62 \\ 
20 & 1 & 20 & 20 & 13 & 3 & Manchester City & 71 & 1.3333 & 172.5 & 153 & -194.805 & 353.423 \\ 
16 & 17 & 8 & 8 & 20 & 4 & Arsenal & 68 & 4.6 & 5.7 & 5.7 & 12.971 & 189.004 \\ 
15 & 10 & 13 & 15 & 10 & 5 & Tottenham & 62 & 1.0667 & 22.5 & 22.5 & -6.154 & 167.724 \\ 
18 & 3 & 18 & 18 & 7 & 6 & Liverpool & 58 & 0.84211 & 91.4 & 89.4 & -49.408 & 276.545 \\ 
13 & 12 & 9 & 9 & 15 & 7 & Everton & 54 & 1.4167 & 6.6 & 6.6 & -5.413 & 103.305 \\ 
8 & 14 & 12 & 12 & 17 & 8 & Fulham & 49 & 2.3333 & 9.3 & 9.3 & 4.406 & 86.369 \\ 
11 & 8 & 17 & 13 & 5 & 9 & Aston Villa & 48 & 0.76471 & 38.2 & 16.7 & -9.744 & 98.564 \\ 
14 & 9 & 14 & 16 & 2 & 10 & Sunderland & 47 & 0.55 & 23 & 23 & -6.234 & 111.154 \\ 
4 & 18 & 10 & 10 & 14 & 11 & West Bromwich & 47 & 1.3571 & 8.58 & 8.58 & 17.049 & 62.777 \\ 
10 & 20 & 6 & 6 & 6 & 12 & Newcastle & 46 & 0.78947 & 4.2 & 4.2 & 32.768 & 92.434 \\ 
6 & 11 & 3 & 2 & 9 & 13 & Stoke & 46 & 1 & 1.9 & 0 & -5.558 & 74.534 \\ 
9 & 4 & 16 & 17 & 4 & 14 & Bolton & 46 & 0.73684 & 27.1 & 27.1 & -26.055 & 87.81 \\ 
7 & 5 & 5 & 5 & 19 & 15 & Blackburn & 43 & 3.3333 & 4 & 4 & -18.615 & 76.173 \\ 
1 & 13 & 4 & 4 & 16 & 16 & Wigan & 42 & 1.6364 & 3.5 & 3.5 & -0.199 & 3.486 \\ 
3 & 15 & 2 & 3 & 8 & 17 & Wolverhampton & 40 & 0.9 & 0.68 & 0.09 & 9.163 & 55.238 \\ 
5 & 7 & 15 & 14 & 3 & 18 & Birmingham & 39 & 0.61905 & 23.2 & 19.2 & -12.342 & 74.383 \\ 
2 & 19 & 1 & 1 & 1 & 19 & Blackpool & 39 & 0.34483 & 0.395 & 0 & 20.783 & 30.893 \\ 
12 & 6 & 7 & 7 & 12 & 20 & West Ham & 33 & 1.2778 & 4.5 & 4.5 & -18.565 & 99.528 \\ 
\hline 
\end{tabular}
}
\caption{Re-rank table for 2010/11 Season. Total Player Spend, Total Foreign Spend, Profits and Expenditure are in millions.}
\label{table:A8}
\end{table}

\begin{table}[ht!]
\centering
\resizebox{\linewidth}{!}{%
\begin{tabular}{|c|c|c|c|c|c|c|c|c|c|c|c|c|}
\hline
 Expense & Profit & Total & Foreign & Foreign: & Position & Team & Pts & Ratio & Total  & Foreign & Profits & Exp\\ 
 (\pounds m) & (\pounds m) & spend (\pounds m) & spend (\pounds m) & British  &  & &  &  & spend (\pounds m) & spend (\pounds m) & (\pounds m) & (\pounds m) \\
 %\midrule
\hline 
20 & 2 & 12 & 14 & 17 & 1 & Chelsea & 86 & 2.4444 & 21.8 & 21.8 & -70.437 & 257.727 \\ 
17 & 18 & 13 & 15 & 10 & 2 & Manchester United & 85 & 1.1111 & 22 & 22 & 13.544 & 191.568 \\ 
16 & 20 & 10 & 10 & 18 & 3 & Arsenal & 75 & 3.25 & 11.2 & 11.2 & 92.32 & 179.496 \\ 
15 & 10 & 14 & 11 & 6 & 4 & Tottenham & 70 & 0.8 & 22.6 & 12.1 & -5.163 & 134.517 \\ 
19 & 1 & 20 & 20 & 14 & 5 & Manchester City & 67 & 1.5833 & 89.3 & 61.8 & -117.793 & 253.801 \\ 
14 & 4 & 18 & 13 & 3 & 6 & Aston Villa & 64 & 0.66667 & 35.6 & 21.6 & -27.712 & 117.198 \\ 
18 & 7 & 19 & 17 & 19 & 7 & Liverpool & 63 & 3.4286 & 44.8 & 23.3 & -19.935 & 227.683 \\ 
13 & 12 & 15 & 16 & 7 & 8 & Everton & 61 & 0.9375 & 23 & 23 & -3.093 & 101.26 \\ 
5 & 16 & 5 & 1 & 1 & 9 & Birmingham & 50 & 0.52632 & 3.4 & 0 & 0.199 & 56.515 \\ 
8 & 13 & 2 & 4 & 20 & 10 & Blackburn & 50 & 3.5 & 2.3 & 2.3 & -1.896 & 70.425 \\ 
7 & 11 & 11 & 12 & 8 & 11 & Stoke & 47 & 1 & 18 & 14.5 & -4.517 & 66.532 \\ 
11 & 8 & 7 & 7 & 13 & 12 & Fulham & 46 & 1.5 & 4.7 & 4.7 & -16.942 & 97.02 \\ 
12 & 5 & 17 & 18 & 4 & 13 & Sunderland & 44 & 0.6875 & 33.3 & 26.3 & -26.179 & 97.149 \\ 
9 & 3 & 16 & 19 & 12 & 14 & Bolton & 39 & 1.4615 & 27.1 & 27.1 & -35.443 & 90.461 \\ 
4 & 19 & 9 & 9 & 5 & 15 & Wolverhampton & 38 & 0.77778 & 10.5 & 10.5 & 16.29 & 44.354 \\ 
1 & 15 & 4 & 6 & 16 & 16 & Wigan & 36 & 2 & 2.9 & 2.9 & 0.075 & 3.677 \\ 
10 & 6 & 3 & 5 & 11 & 17 & West Ham & 35 & 1.1429 & 2.5 & 2.5 & -21.485 & 94.262 \\ 
3 & 17 & 6 & 2 & 2 & 18 & Burnley & 30 & 0.52632 & 3.5 & 0 & 10.247 & 40.372 \\ 
6 & 9 & 8 & 8 & 9 & 19 & Hull & 30 & 1 & 5.8 & 5.8 & -6.831 & 58.154 \\ 
2 & 14 & 1 & 3 & 15 & 20 & Portsmouth & 19 & 1.8 & 2 & 0 & 0 & 5.4 \\ 
\hline 
\end{tabular}
}
\caption{Re-rank table for 2009/10 Season. Total Player Spend, Total Foreign Spend, Profits and Expenditure are in millions.}
\label{table:A9}
\end{table}
\newpage
\clearpage
%%%%%%%%%%%%%%%%%%%%%%%%%%%%%%%%%%
\section*{\large Appendix 2: Principal Component Analysis (PCA)}
% \textbf{PCA for Each Season}

\begin{table}[ht]
\centering
\resizebox{\linewidth}{!}{%
\begin{tabular}{|c|c|c|c|c|c|c|c|c|c|c|c|c|}
\hline
Variable & PCA 1 & PCA 2 & PCA 3 & PCA 4 & PCA 5 & PCA 6 \\ 
\hline 
Pts & 0.117 & -0.0283 & 0.316 & 0.736 & -0.586 & 0.00643 \\ 
Ratio & 0.00363 & 0.00345 & 0.00765 & -0.0133 & -0.00104 & 0.999 \\ 
Player Spend & 0.315 & 0.533 & 0.361 & -0.537 & -0.443 & -0.0134 \\ 
Foreign Spend & 0.276 & 0.551 & 0.277 & 0.367 & 0.640 & 0.000523 \\ 
Profits & -0.0590 & -0.512 & 0.808 & -0.181 & 0.221 & -0.00639 \\ 
Exp & 0.898 & -0.386 & -0.200 & -0.0322 & 0.0498 & -0.000767 \\ 
\hline 
\end{tabular}
}
\caption{PCA for 2016/17 Season}
\label{table:A11}
\end{table}

\begin{table}[ht!]
\centering
\resizebox{\linewidth}{!}{%
\begin{tabular}{|c|c|c|c|c|c|c|c|c|c|c|c|c|}
\hline
Variable & PCA 1 & PCA 2 & PCA 3 & PCA 4 & PCA 5 & PCA 6 \\ 
\hline 
Pts & 0.0420 & 0.00439 & 0.221 & 0.958 & -0.178 & 0.000951 \\ 
Ratio & 0.00284 & -0.00357 & 0.00645 & -0.00468 & -0.0113 & 0.999 \\ 
Player Spend & 0.292 & 0.741 & 0.168 & -0.159 & -0.560 & -0.00631 \\ 
Foreign Spend & 0.239 & 0.554 & -0.152 & 0.164 & 0.766 & 0.0117 \\ 
Profits & -0.0888 & -0.0162 & 0.946 & -0.165 & 0.263 & -0.00372 \\ 
Exp & 0.921 & -0.380 & 0.0675 & -0.0520 & 0.0119 & -0.00452 \\ 
\hline 
\end{tabular}
}
\caption{PCA for 2015/16 Season}
\label{table:A12}
\end{table}

% {\color{red}{Use Figure 4.1 or the most "attractive one" in the main results' section and complement the discussion under "Conclusion". Figures 4.2, 4.3, etc can remain in the appendix.}}
\begin{table}[ht!]
\centering
\resizebox{\linewidth}{!}{%
\begin{tabular}{|c|c|c|c|c|c|c|c|c|c|c|c|c|}
\hline
Variable & PCA 1 & PCA 2 & PCA 3 & PCA 4 & PCA 5 & PCA 6 \\ 
\hline 
Pts & 0.116 & 0.0271 & -0.0653 & 0.985 & -0.103 & -0.0186 \\ 
Ratio & 0.00468 & 0.00226 & -0.0103 & 0.0232 & 0.0542 & 0.998 \\ 
Player Spend & 0.412 & -0.0254 & 0.631 & -0.0733 & -0.652 & 0.0417 \\ 
Foreign Spend & 0.388 & 0.0323 & 0.531 & 0.0657 & 0.748 & -0.0385 \\ 
Profits & -0.0175 & 0.999 & 0.00906 & -0.0288 & -0.0380 & 0.000639 \\ 
Exp & 0.816 & 0.0149 & -0.562 & -0.136 & -0.0136 & -0.00576 \\ 
\hline 
\end{tabular}
}
\caption{PCA for 2014/15 Season}
\label{table:A13}
\end{table}

\begin{table}[ht!]
\centering
\resizebox{\linewidth}{!}{%
\begin{tabular}{|c|c|c|c|c|c|c|c|c|c|c|c|c|}
\hline
Variable & PCA 1 & PCA 2 & PCA 3 & PCA 4 & PCA 5 & PCA 6 \\ 
\hline 
Pts & 0.117 & -0.0283 & 0.316 & 0.736 & -0.586 & 0.00643 \\ 
Ratio & 0.00363 & 0.00345 & 0.00765 & -0.0133 & -0.00104 & 0.999 \\ 
Player Spend & 0.315 & 0.533 & 0.361 & -0.537 & -0.443 & -0.0134 \\ 
Foreign Spend & 0.276 & 0.551 & 0.277 & 0.367 & 0.640 & 0.000523 \\ 
Profits & -0.0590 & -0.512 & 0.808 & -0.181 & 0.221 & -0.00639 \\ 
Exp & 0.898 & -0.386 & -0.200 & -0.0322 & 0.0498 & -0.000767 \\ 
\hline 
\end{tabular}
}
\caption{PCA for 2013/14 Season}
\label{table:A14}
\end{table}

\begin{table}[ht!]
\centering
\resizebox{\linewidth}{!}{%
\begin{tabular}{|c|c|c|c|c|c|c|c|c|c|c|c|c|}
\hline
Variable & PCA 1 & PCA 2 & PCA 3 & PCA 4 & PCA 5 & PCA 6 \\ 
\hline 
Pts & 0.138 & 0.0928 & 0.377 & 0.911 & -0.00759 & 0.0128 \\ 
Ratio & 0.00256 & 0.00331 & 6.5275$e^{-05}$ & -0.0155 & -0.0874 & 0.996 \\ 
Player Spend & 0.174 & 0.636 & -0.214 & 0.00266 & 0.718 & 0.0605 \\ 
Foreign Spend & 0.165 & 0.695 & -0.118 & -0.0516 & -0.685 & -0.0636 \\ 
Profits & -0.175 & 0.248 & 0.878 & -0.361 & 0.0857 & 0.00148 \\ 
Exp & 0.945 & -0.206 & 0.168 & -0.191 & 0.00396 & -0.00438 \\ 
\hline 
\end{tabular}
}
\caption{PCA for 2012/13 Season}
\label{table:A15}
\end{table}

\begin{table}[ht!]
\centering
\resizebox{\linewidth}{!}{%
\begin{tabular}{|c|c|c|c|c|c|c|c|c|c|c|c|c|}
\hline
Variable & PCA 1 & PCA 2 & PCA 3 & PCA 4 & PCA 5 & PCA 6 \\ 
\hline 
Pts & 0.146 & 0.105 & 0.00416 & 0.981 & -0.0766 & 0.00344 \\ 
Ratio & 0.00440 & 0.00245 & 0.00132 & 0.00255 & 0.0891 & 0.996 \\ 
Player Spend & 0.226 & -0.0306 & 0.710 & -0.0851 & -0.658 & 0.0572 \\ 
Foreign Spend & 0.200 & -0.131 & 0.625 & 0.0395 & 0.739 & -0.0676 \\ 
Profits & -0.136 & 0.972 & 0.153 & -0.0778 & 0.0825 & -0.00918 \\ 
Exp & 0.932 & 0.161 & -0.285 & -0.153 & 0.0252 & -0.00600 \\ 
\hline 
\end{tabular}
}
\caption{PCA for 2011/12 Season}
\label{table:A16}
\end{table}

\begin{table}[ht!]
\centering
\resizebox{\linewidth}{!}{%
\begin{tabular}{|c|c|c|c|c|c|c|c|c|c|c|c|c|}
\hline
Variable & PCA 1 & PCA 2 & PCA 3 & PCA 4 & PCA 5 & PCA 6 \\ 
\hline 
Pts & 0.0806 & 0.183 & -0.0963 & 0.969 & 0.107 & -0.0275 \\ 
Ratio & 0.00109 & 0.00916 & -0.0259 & 0.0176 & 0.0571 & 0.998 \\ 
Player Spend & 0.367 & -0.374 & 0.485 & 0.165 & -0.679 & 0.0515 \\ 
Foreign Spend & 0.343 & -0.301 & 0.521 & -0.000471 & 0.721 & -0.0253 \\ 
Profits & -0.363 & 0.630 & 0.684 & -0.0138 & -0.0574 & 0.0159 \\ 
Exp & 0.781 & 0.582 & -0.129 & -0.184 & -0.0357 & -0.004245 \\ 
\hline 
\end{tabular}
}
\caption{PCA for 2010/11 Season}
\label{table:A17}
\end{table}

\begin{table}[ht!]
\centering
\resizebox{\linewidth}{!}{%
\begin{tabular}{|c|c|c|c|c|c|c|c|c|c|c|c|c|}
\hline
Variable & PCA 1 & PCA 2 & PCA 3 & PCA 4 & PCA 5 & PCA 6 \\ 
\hline 
Pts & 0.188 & 0.192 & -0.0375 & 0.958 & -0.0945 & 0.0258 \\ 
Ratio & 0.00377 & 0.00868 & -0.0111 & -0.0317 & -0.0184 & 0.999 \\ 
Player Spend & 0.195 & -0.220 & 0.773 & -0.0194 & -0.562 & -0.00118 \\ 
Foreign Spend & 0.138 & -0.145 & 0.524 & 0.103 & 0.821 & 0.0249 \\ 
Profits & -0.243 & 0.907 & 0.323 & -0.120 & 0.00956 & -0.00699 \\ 
Exp & 0.921 & 0.268 & -0.149 & -0.238 & 0.0177 & -0.0147 \\ 
\hline 
\end{tabular}
}
\caption{PCA for 2009/10 Season}
\label{table:A18}
\end{table}

\newpage
%\addcontentsline{toc}{chapter}{Appendix}
\section*{\large Appendix 3: Indicators comparing footballing performance against sectorial investments on players, including foreign players}
\setcounter{figure}{0}
\renewcommand{\thefigure}{F\arabic{figure}}
 \textbf{}
\begin{figure}[H]
\begin{subfigure}[b]{0.48\textwidth}
	\centering
	\includegraphics[width=\textwidth]{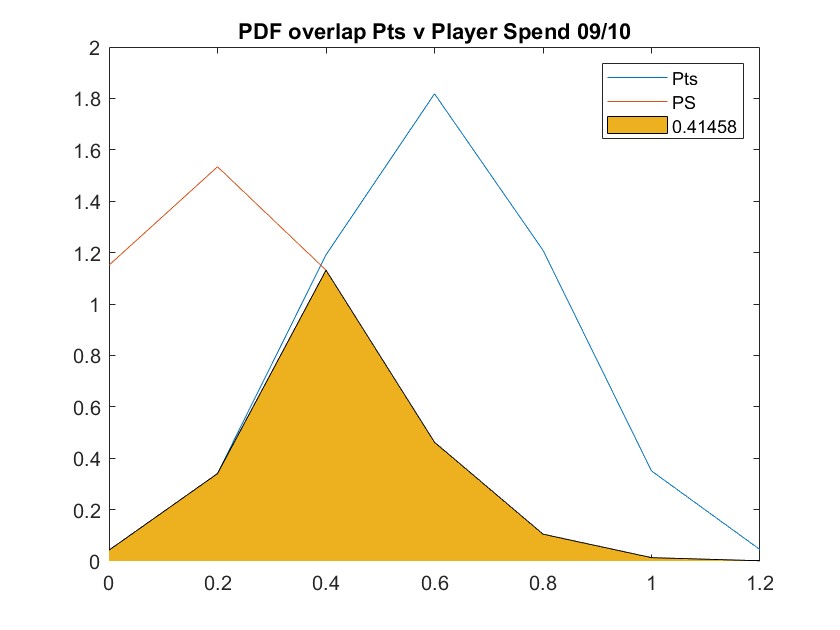}
	\caption{\centering PDF overlaps of foreign spend to profit = 41.45\%}
	\centering
	\label{label:file_name}
\end{subfigure}
\hfill
\begin{subfigure}[b]{0.48\textwidth}
	\centering
	\includegraphics[width=\textwidth]{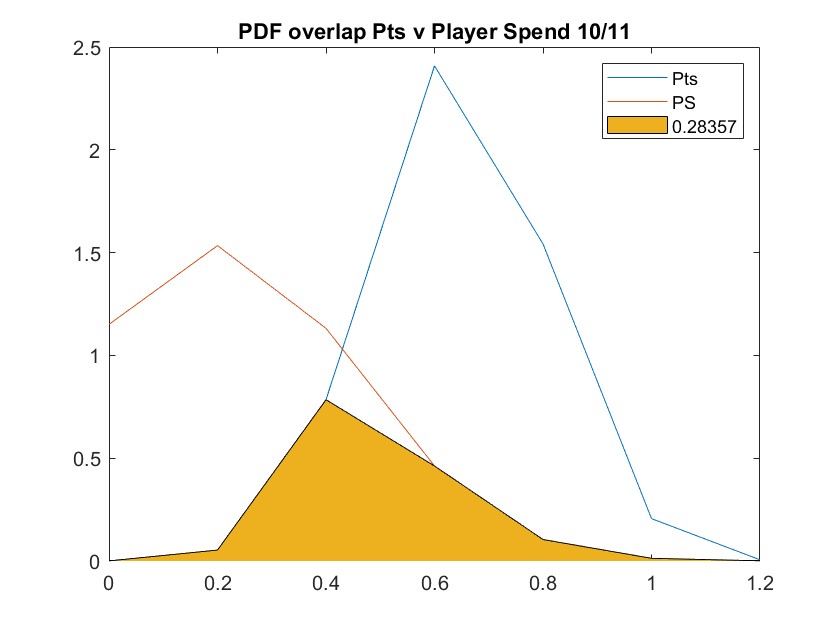}
	\caption{\centering PDF overlaps of foreign spend to profit = 28.36\%}
	\centering
	\label{label:file_name}
\end{subfigure}
\hfill
\begin{subfigure}[b]{0.48\textwidth}
	\centering
	\includegraphics[width=\textwidth]{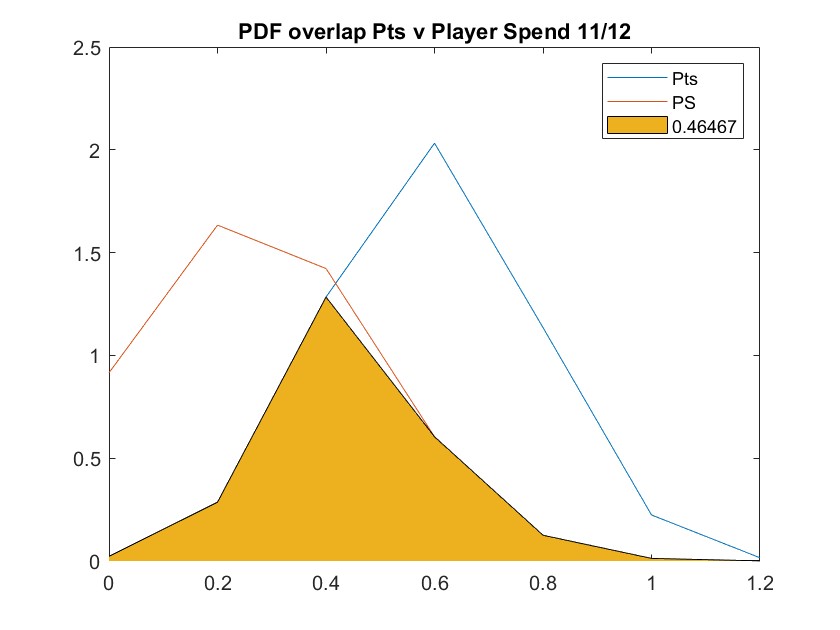}
	\caption{\centering PDF overlaps of foreign spend to profit = 46.47\%}
	\centering
	%\label{label:file_name}
\end{subfigure}
\hfill
\begin{subfigure}[b]{0.48\textwidth}
	\centering
	\includegraphics[width=\textwidth]{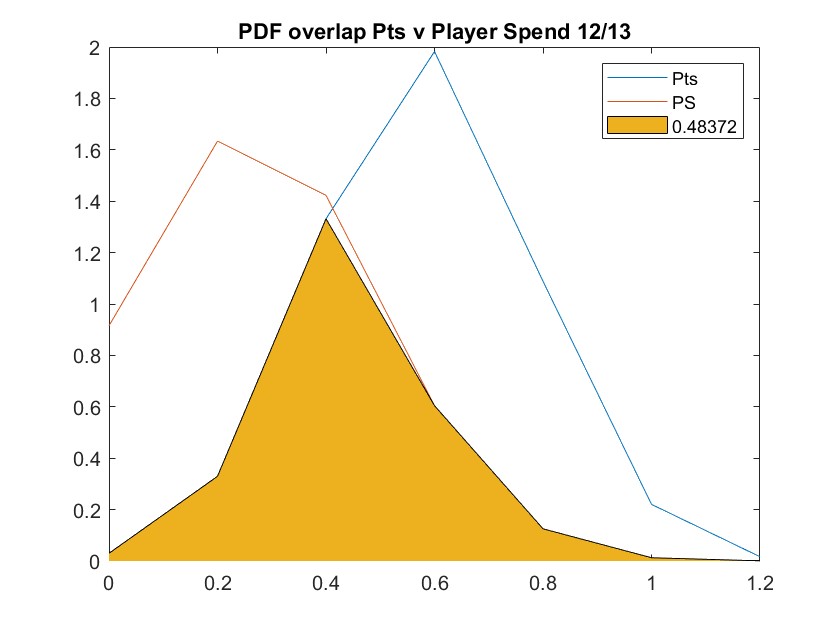}
	\caption{\centering PDF overlaps of foreign spend to profit = 48.37\%}
	\centering
	%\label{label:file_name}
	\end{subfigure}
	\hfill
	\begin{subfigure}[b]{0.48\textwidth}
	\centering
	\includegraphics[width=\textwidth]{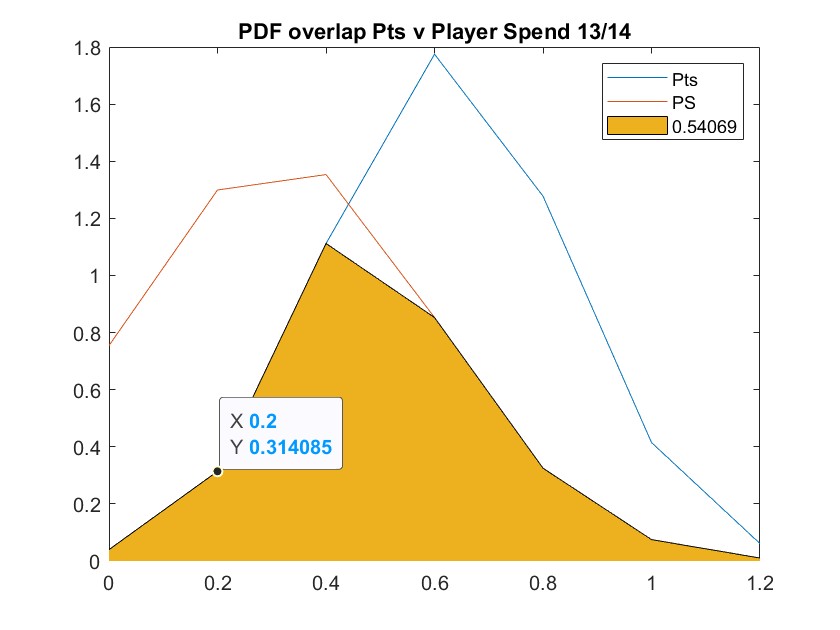}
	\caption{PDF overlaps of foreign spend to profit = 54.07\%}
	\centering
	\label{label:file_name}
\end{subfigure}
\hfill
\begin{subfigure}[b]{0.48\textwidth}
	\centering
	\includegraphics[width=\textwidth]{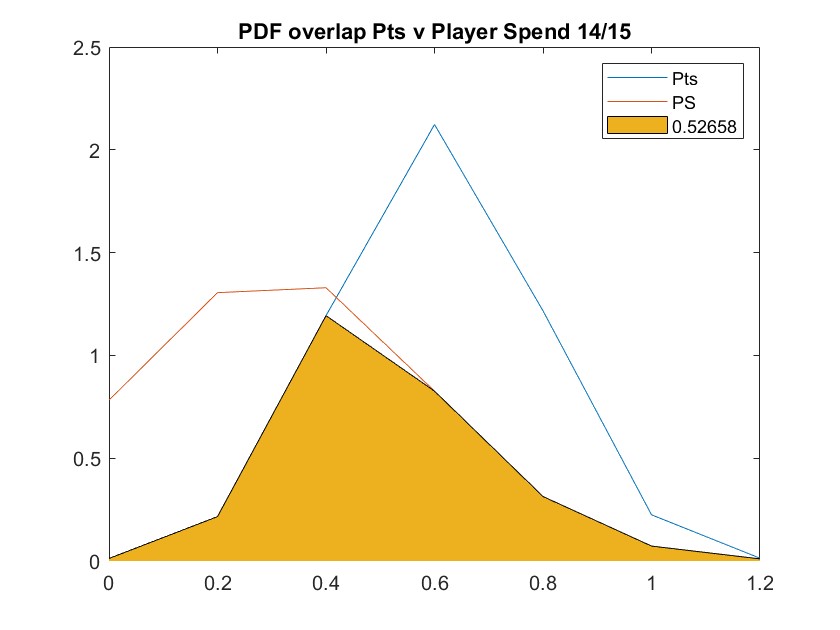}
	\caption{\centering PDF overlaps of foreign spend to profit = 52.66\%}
	\centering
	%\label{label:file_name}
\end{subfigure}
\hfill
\begin{subfigure}[b]{0.48\textwidth}
	\centering
	\includegraphics[width=\textwidth]{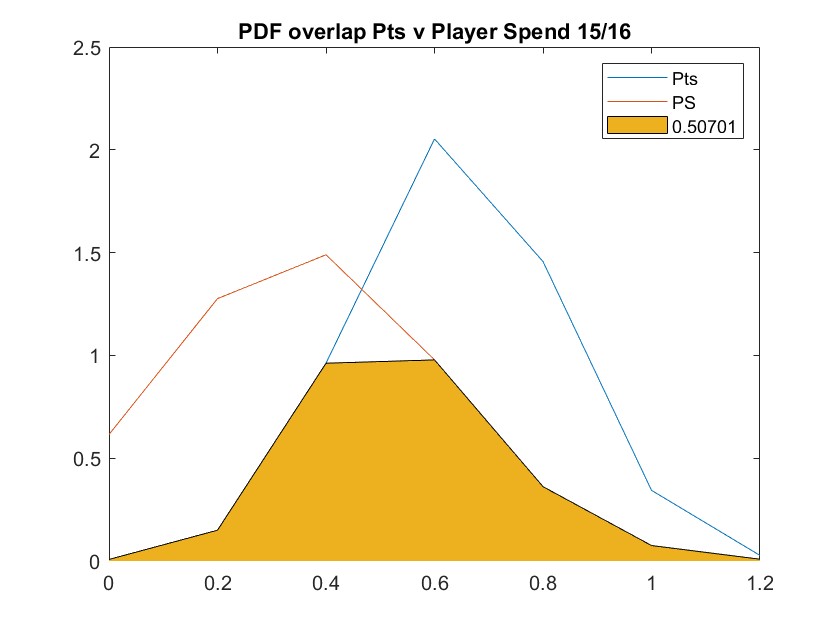}
	\caption{\centering PDF overlaps of foreign spend to profit = 50.709\%}
	\centering
	%\label{label:file_name}
\end{subfigure}
\hfill
\begin{subfigure}[b]{0.48\textwidth}
	\centering
	\includegraphics[width=\textwidth]{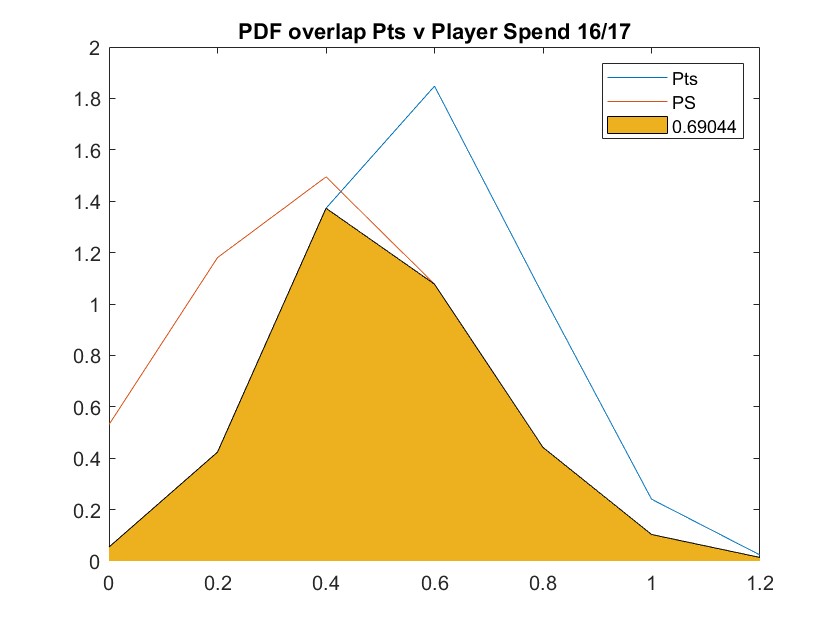}
	\caption{\centering PDF overlaps of foreign spend to profit = 63.75\%}
	\centering
	%\label{label:file_name}
\end{subfigure}
\caption{Percentage of spends on foreign players that translated into economic profit.}
\label{figure:F1}
\end{figure}

\begin{figure}[H]
\begin{subfigure}[b]{0.48\textwidth}
	\centering
	\includegraphics[width=\textwidth]{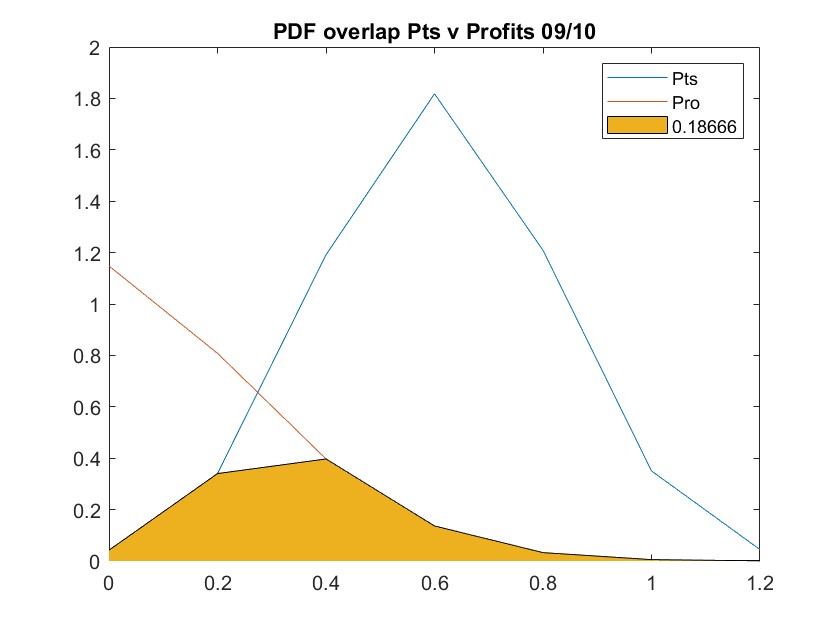}
	\caption{\centering PDF overlaps of points to profit =18.67\%}
	\centering
	\label{label:file_name}
\end{subfigure}
\hfill
\begin{subfigure}[b]{0.48\textwidth}
	\centering
	\includegraphics[width=\textwidth]{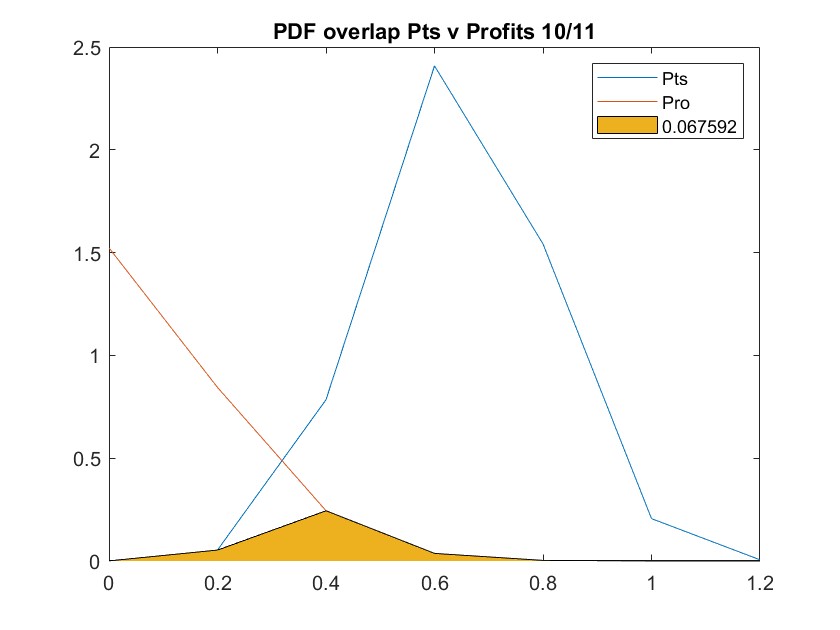}
	\caption{\centering PDF overlaps of points to profit = 6.76\%}
	\centering
	\label{label:file_name}
\end{subfigure}
\hfill
\begin{subfigure}[b]{0.48\textwidth}
	\centering
	\includegraphics[width=\textwidth]{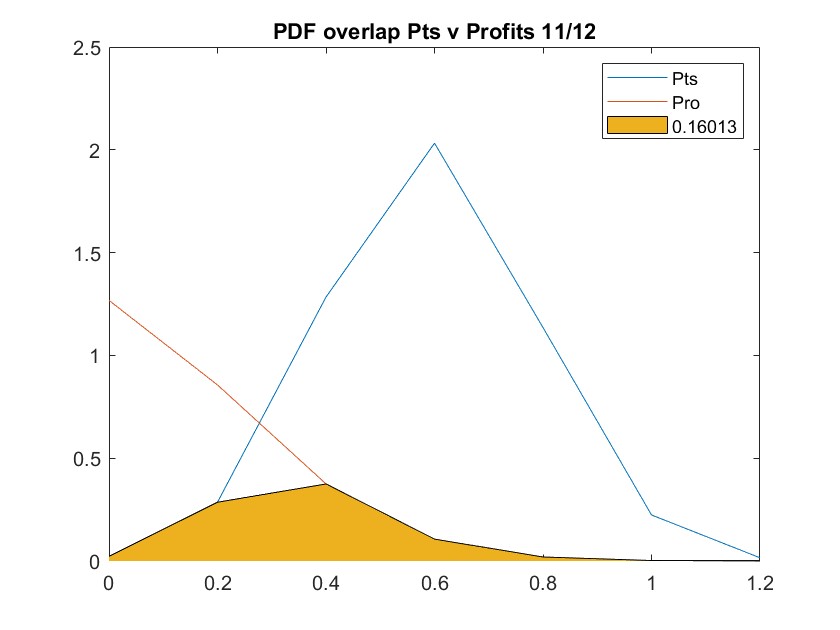}
	\caption{\centering PDF overlaps of points to profit = 16.01\%}
	\centering
	%\label{label:file_name}
\end{subfigure}
\hfill
\begin{subfigure}[b]{0.48\textwidth}
	\centering
	\includegraphics[width=\textwidth]{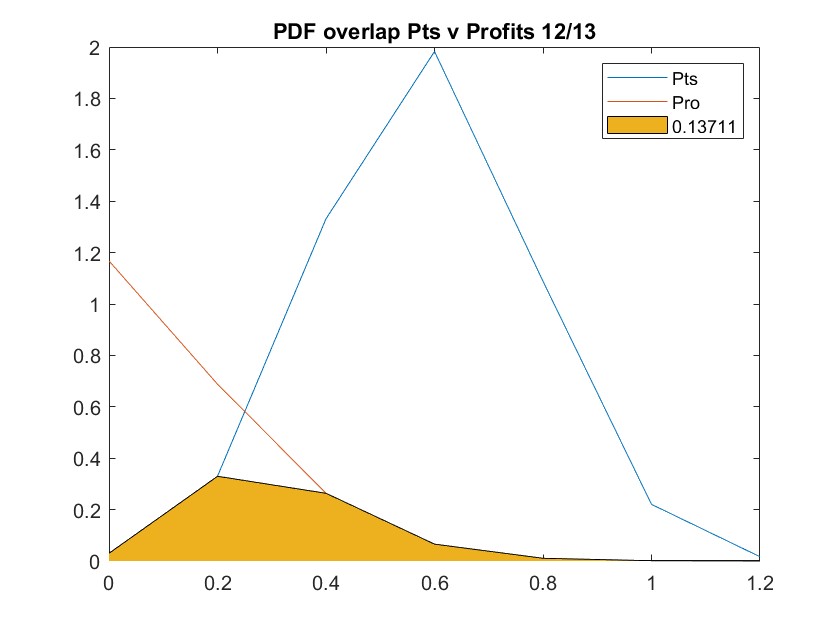}
	\caption{\centering PDF overlaps of points to profit = 13.71\%}
	\centering
	%\label{label:file_name}
	\end{subfigure}
	\hfill
	\begin{subfigure}[b]{0.48\textwidth}
	\centering
	\includegraphics[width=\textwidth]{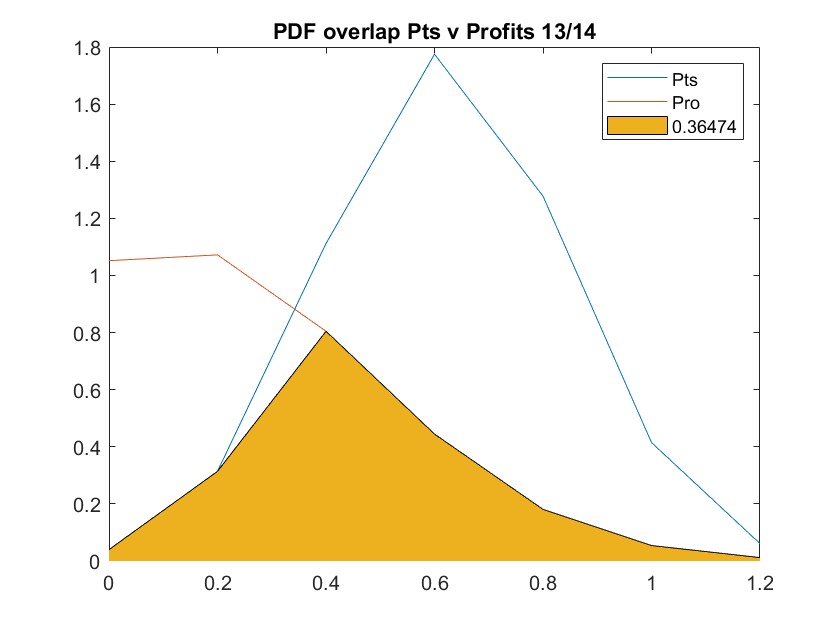}
	\caption{PDF overlaps of points to profit = 36.47\%}
	\centering
	\label{label:file_name}
\end{subfigure}
\hfill
\begin{subfigure}[b]{0.48\textwidth}
	\centering
	\includegraphics[width=\textwidth]{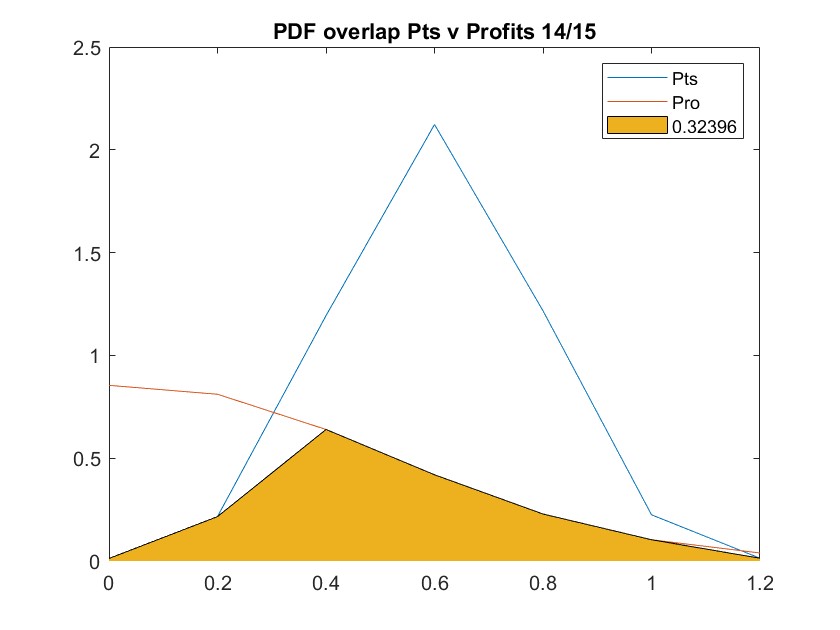}
	\caption{\centering PDF overlaps of points to profit = 32.40\%}
	\centering
	%\label{label:file_name}
\end{subfigure}
\hfill
\begin{subfigure}[b]{0.48\textwidth}
	\centering
	\includegraphics[width=\textwidth]{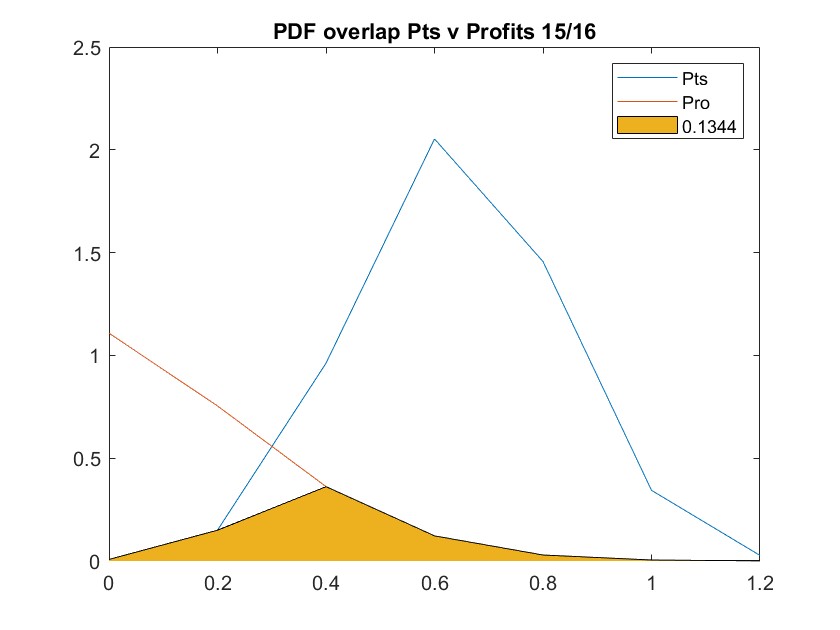}
	\caption{\centering PDF overlaps of points to profit = 13.44\%}
	\centering
	%\label{label:file_name}
\end{subfigure}
\hfill
\begin{subfigure}[b]{0.48\textwidth}
	\centering
	\includegraphics[width=\textwidth]{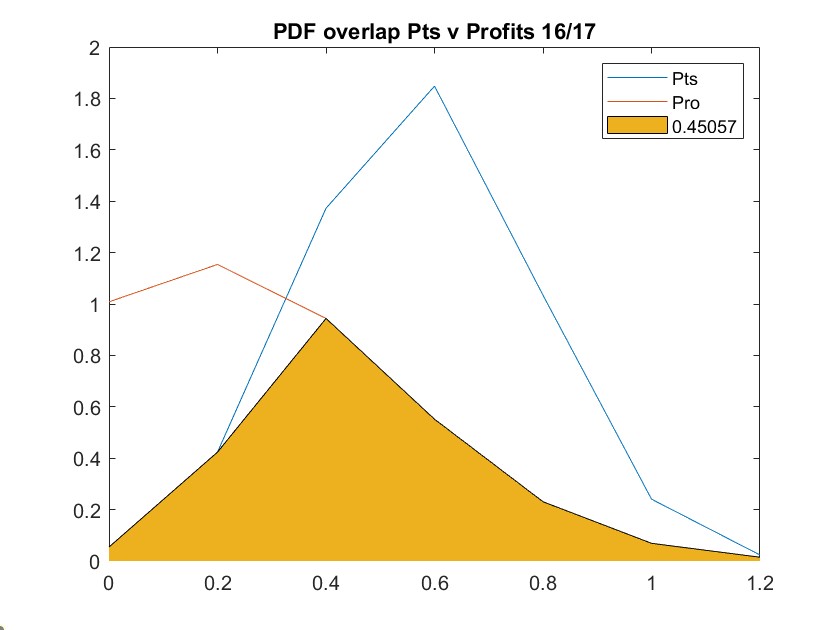}
	\caption{\centering PDF overlaps of points to profit = 45.06\%}
	\centering
	%\label{label:file_name}
\end{subfigure}
\caption{Correlation between on-field performance measured as points scored against economic profits.}
\label{figure:F2}
\end{figure}

\begin{figure}[H]
\begin{subfigure}[b]{0.48\textwidth}
	\centering
	\includegraphics[width=\textwidth]{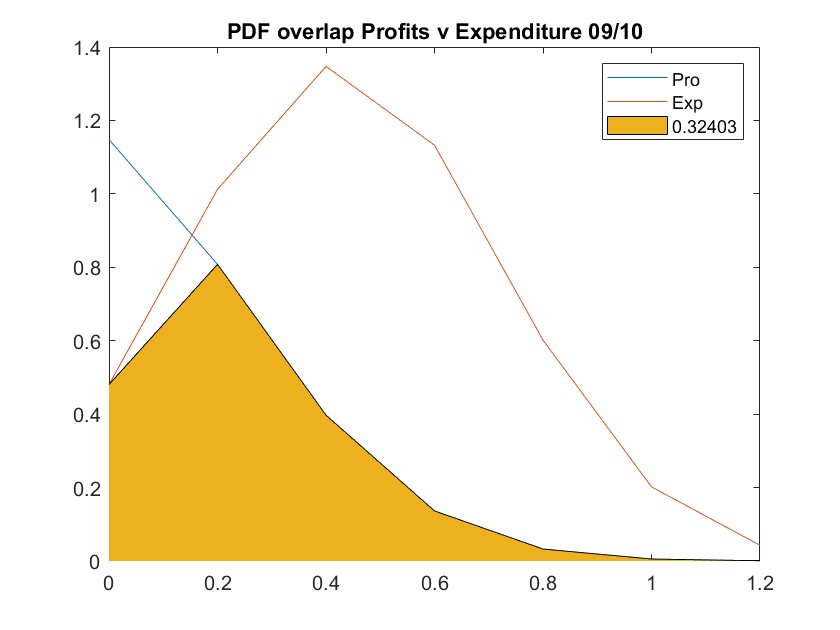}
	\caption{\centering PDF overlaps of profit against expenditure =32.40\%}
	\centering
	\label{label:file_name}
\end{subfigure}
\hfill
\begin{subfigure}[b]{0.48\textwidth}
	\centering
	\includegraphics[width=\textwidth]{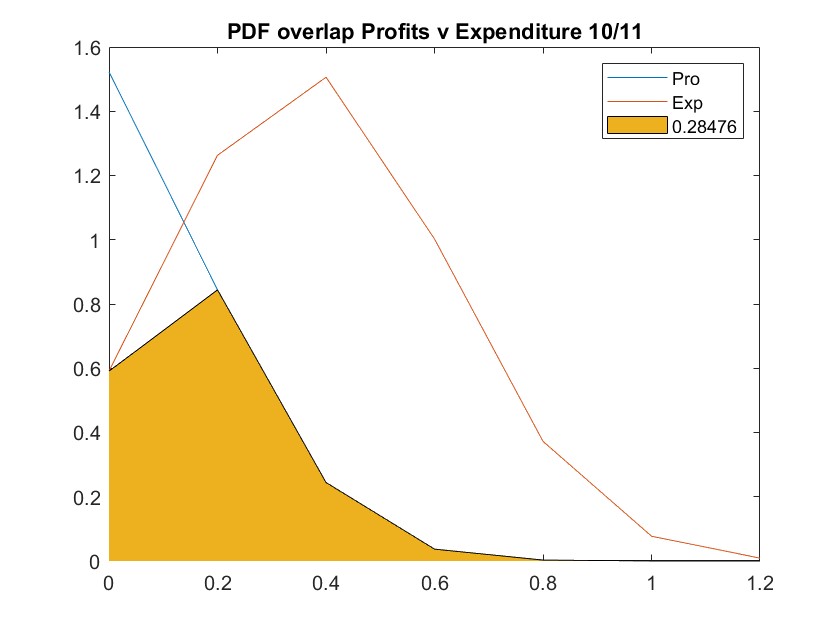}
	\caption{\centering PDF overlaps of profit against expenditure = 28.47\%}
	\centering
	\label{label:file_name}
\end{subfigure}
\hfill
\begin{subfigure}[b]{0.48\textwidth}
	\centering
	\includegraphics[width=\textwidth]{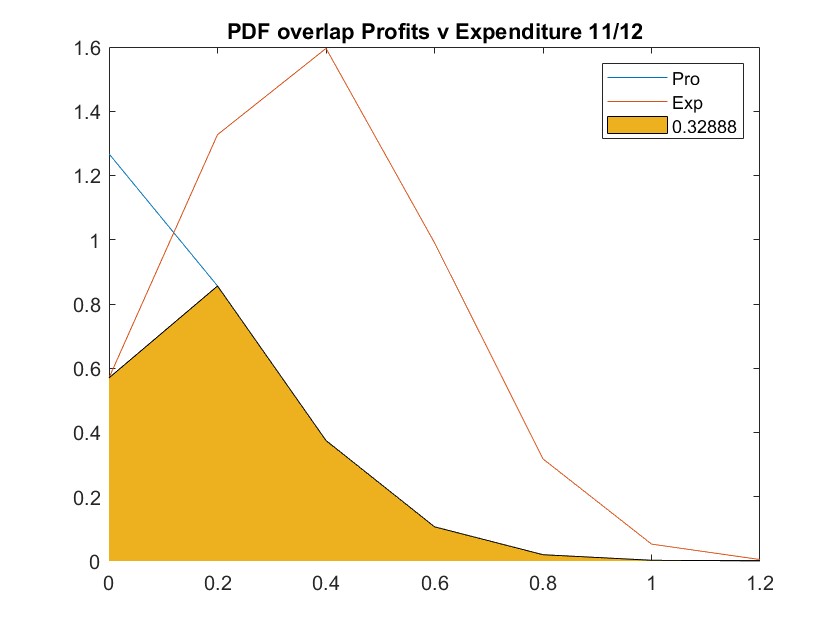}
	\caption{\centering PDF overlaps of profit against expenditure = 32.89\%}
	\centering
	%\label{label:file_name}
\end{subfigure}
\hfill
\begin{subfigure}[b]{0.48\textwidth}
	\centering
	\includegraphics[width=\textwidth]{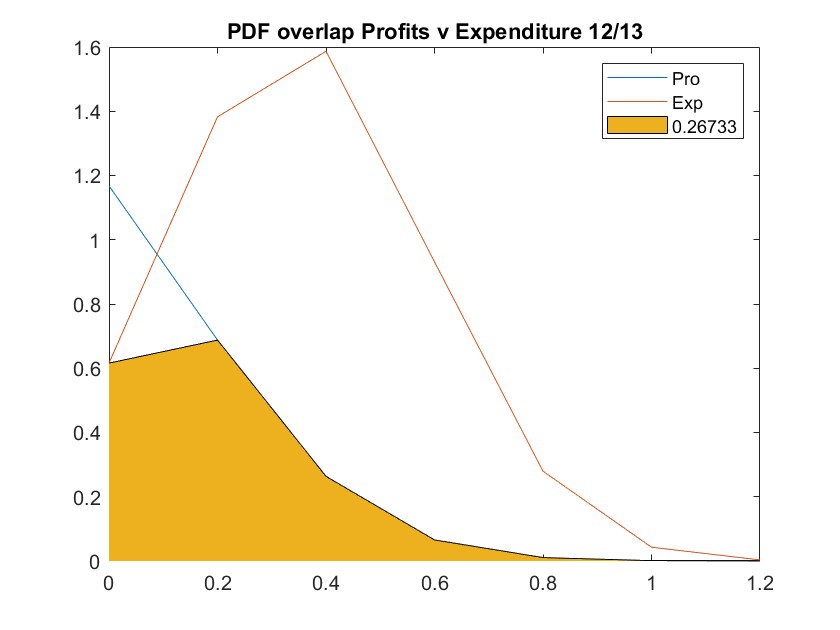}
	\caption{\centering PDF overlaps of profit against expenditure = 26.73\%}
	\centering
	%\label{label:file_name}
	\end{subfigure}
	\hfill
	\begin{subfigure}[b]{0.48\textwidth}
	\centering
	\includegraphics[width=\textwidth]{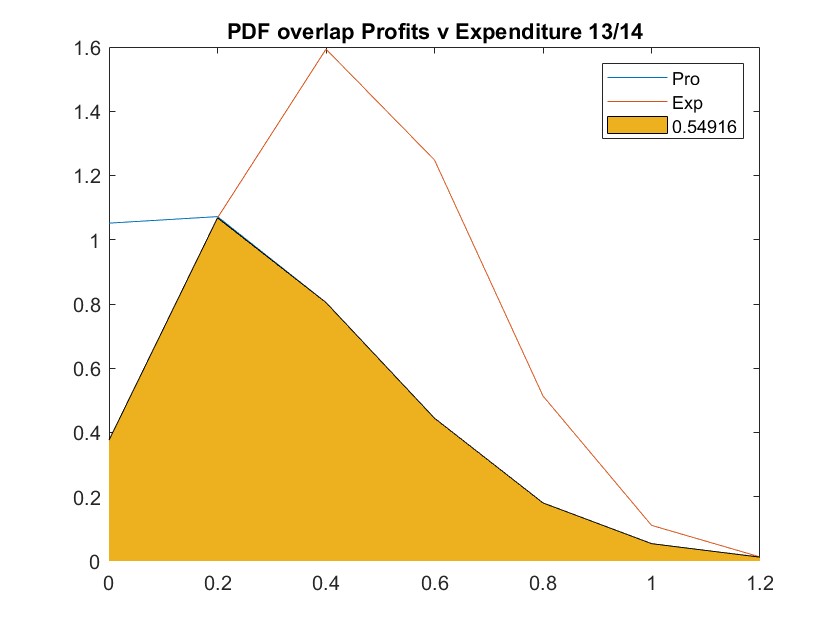}
	\caption{PDF overlaps of profit against expenditure = 54.92\%}
	\centering
	\label{label:file_name}
\end{subfigure}
\hfill
\begin{subfigure}[b]{0.48\textwidth}
	\centering
	\includegraphics[width=\textwidth]{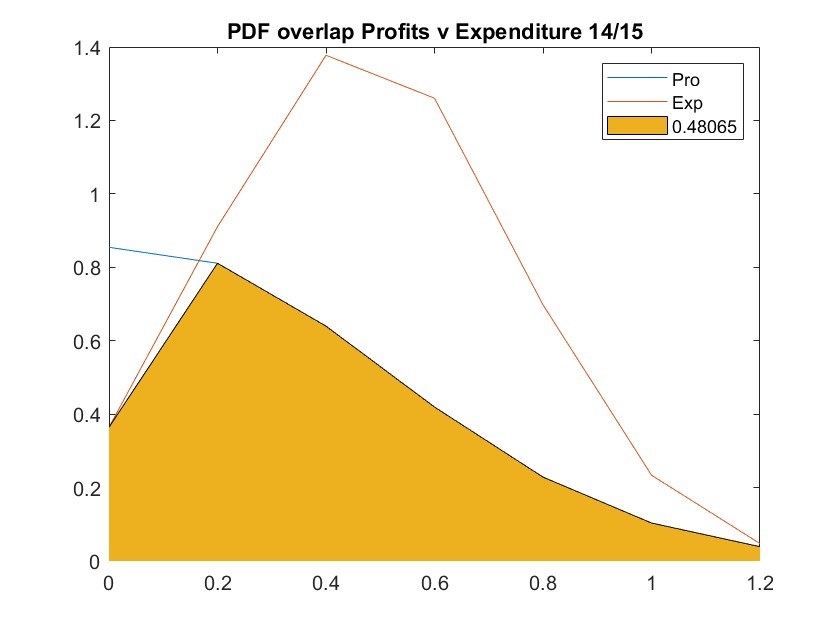}
	\caption{\centering PDF overlaps of profit against expenditure = 48.06\%}
	\centering
	%\label{label:file_name}
\end{subfigure}
\hfill
\begin{subfigure}[b]{0.48\textwidth}
	\centering
	\includegraphics[width=\textwidth]{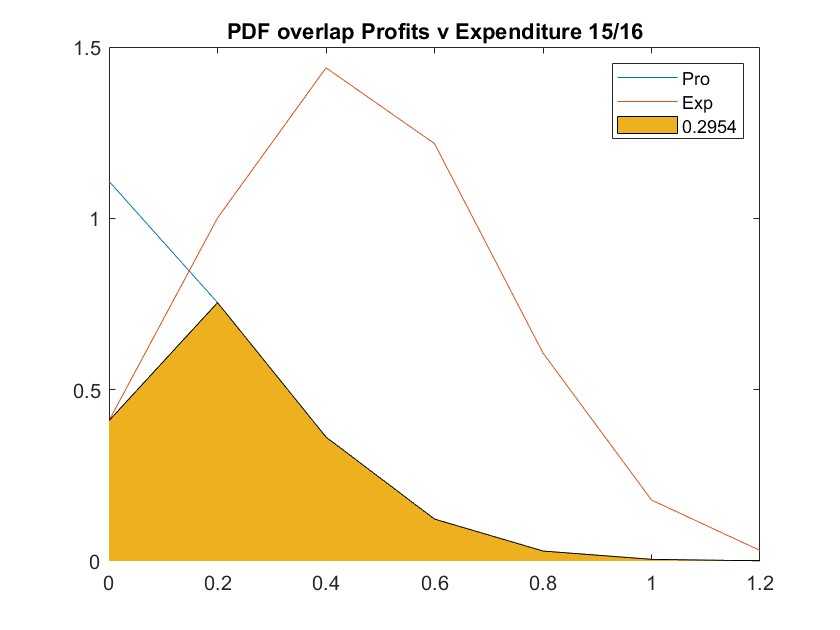}
	\caption{\centering PDF overlaps of profit against expenditure = 29.54\%}
	\centering
	%\label{label:file_name}
\end{subfigure}
\hfill
\begin{subfigure}[b]{0.48\textwidth}
	\centering
	\includegraphics[width=\textwidth]{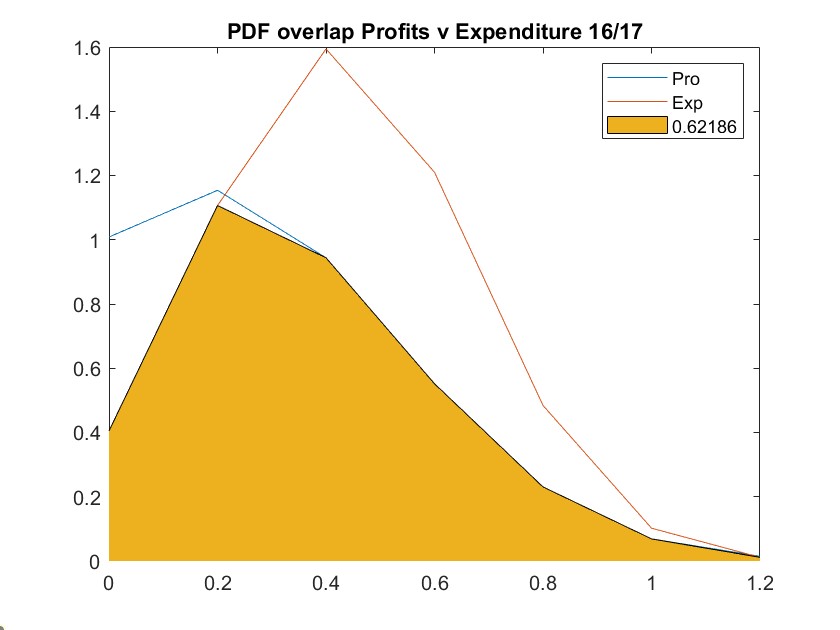}
	\caption{\centering PDF overlaps of profit against expenditure = 62.19\%}
	\centering
	%\label{label:file_name}
\end{subfigure}
\caption{Correlation between economic profit against total expenditure.}
\label{figure:F3}
\end{figure}

\end{document}